\documentclass[%
 reprint,
superscriptaddress,
nofootinbib,
 amsmath,amssymb,
 aps,
 prd,
]{revtex4-2}

\usepackage{graphicx}
\usepackage{xcolor}
\usepackage{pifont}
\usepackage{dcolumn}
\usepackage{bm}

\providecommand{\blip}{\texttt{BLIP} }

\usepackage{hyperref}

\begin{document}

\preprint{APS/123-QED}

\title{The Gravitational Wave Foreground from Galactic Cataclysmic Variables in LISA}

\author{Levi Schult}
\email[Corresponding author: ]{levi.s.schult@vanderbilt.edu}
\affiliation{Department of Physics and Astronomy, Vanderbilt University, 2301 Vanderbilt Place, Nashville, TN 37235, USA}

\author{Alexander Criswell}%
\affiliation{Department of Physics and Astronomy, Vanderbilt University, 2301 Vanderbilt Place, Nashville, TN 37235, USA}
\affiliation{Department of Life and Physical Sciences, Fisk University, 1000 17th Avenue N., Nashville, 37208, TN, USA}

\author{Stephen R. Taylor}
\affiliation{Department of Physics and Astronomy, Vanderbilt University, 2301 Vanderbilt Place, Nashville, TN 37235, USA}

\author{Katelyn Breivik}
\affiliation{McWilliams Center for Cosmology and Astrophysics, Department of Physics,
Carnegie Mellon University, Pittsburgh, PA 15213, USA}%

\date{\today}

\begin{abstract}
Electromagnetic observations have been the primary medium for studying cataclysmic variable (CV) systems, capitalizing on the invaluable opportunity they present to study binary stellar evolution and accretion. The launch of the Laser Interferometer Space Antenna (LISA) mission, however, will enable a multi-messenger perspective of the population. We extend previous work to explore the gravitational wave (GW) spectra sourced by a local and Galactic population of CVs. 
Using the Bayesian LISA Inference Package (BLIP), we simulate LISA data containing these CV populations 
to understand their impact and our ability to jointly model CV and double white dwarf (DWD) GW spectra. While we find that a local CV population has negligible effect on the LISA datastream, a Galactic CV population will have a signal-to-noise ratio of $\approx76$, depending on the characteristics of the DWD population present, and will bias existing DWD spectral models if ignored. 
Current approaches struggle to accurately recover both populations' respective spectra, calling for the creation of multi-component spectral models that include GW contributions from both CVs and DWDs.  
These results point toward a promising future for the study of CVs with LISA.
\end{abstract}

\maketitle


\section{Introduction}
\label{sec:Intro}
The Laser Interferometer Space Antenna (LISA) mission is a space-based gravitational wave (GW) detector that will launch in the 2030s and operate in the millihertz (mHz) frequency regime \citep{colpi_lisa_2024}.  LISA will capture GWs from the rich variety of sources that exist in this frequency band to probe unexplored areas of astrophysics. These sources include massive black hole mergers, extreme mass ratio inspirals, stellar-mass black hole binaries, potential cosmological sources, and galactic stellar binaries (GBs) \citep{amaro-seoane_astrophysics_2023, colpi_lisa_2024}. The resultant data will contain omnipresent GW signals and noise, requiring a ``global fit'' data analysis strategy, in which everything will be modeled simultaneously \citep{colpi_lisa_2024, 2025PhRvD.111b4060K, littenberg_prototype_2023}. 

The Galactic population of stellar binaries is primarily composed of double white dwarf (DWD) systems. While there will be tens of thousands of resolved DWD systems, approximately 30 million DWDs will be unresolved, creating a stochastic GW foreground (GWF) 
that will be greater than LISA's instrumental noise below 1-3 mHz, known as the Galactic foreground \citep{colpi_lisa_2024, amaro-seoane_astrophysics_2023}. 
Accurate modeling of this foreground is critical to extract answers to understand DWDs.
Details of the spectrum promise to reveal insights around binary stellar evolution and its anisotropies can be harnessed to study Milky Way (MW) structure \citep{2020ApJ...901....4B, korol_observationally_2022, amaro-seoane_astrophysics_2023}.
Additionally, due to this foreground surpassing the LISA instrumental noise and the global nature of the data analysis scheme, the study of other sources within LISA data will be limited by the noise of the foreground \citep{colpi_lisa_2024, littenberg_prototype_2023, 2025PhRvD.111b4060K}. 
For these reasons, accurately characterizing the GWF arising from WD binary systems in the MW is of high importance.
 
DWDs are not the only population of GB that will contribute to the Galactic foreground, however. 
This work adds to a growing body examining the impact of cataclysmic variables (CVs) which consist of a WD primary accreting matter from a main sequence star \citep{warner2003cataclysmic, Scaringi2023}.

The study of CVs promises to reveal details of accretion physics and binary stellar evolution. This knowledge will also improve our understanding of the dynamics present in other types of binary systems \citep{tauris_physics_2022, knigge_evolution_2011}. CVs may in some cases be progenitors for Type Ia supernovae through the single degenerate pathway \citep{tauris_physics_2022}. 
Some of the open questions regarding CVs include the processes that remove angular momentum from the system \citep{tauris_physics_2022, schreiber_three_2016, zorotovic_cataclysmic_2020, metzger_transients_2021}, their role as a progenitor for another class of mass-transferring stellar binary (AM Canum Venaticorum systems) \citep{belloni_reversing_2023}, and their true space density \citep{Pala2020, rodriguez_cataclysmic_2025}. Questions like these can only be answered by studying the CV population as a whole, and GWs will grant a perspective of CVs unhindered by biases that inhibit EM efforts \citep{rodriguez_cataclysmic_2025, Scaringi2023}.

Here, we summarize CV evolution broadly, but we direct the interested reader to \citet{knigge_evolution_2011} and \citet{zorotovic_cataclysmic_2020} for further review. 
The observed population exists over the $0.05-0.45$~mHz GW frequency range (orbital periods of $\approx 1-6$ hrs) and have several unique features as a population.
In a CV system, mass transfer occurs due to the main sequence secondary filling its Roche lobe \citep{warner2003cataclysmic}. At low GW frequencies (high orbital period), angular momentum loss in the system is driven by the donor star's magnetized stellar winds in a process known as magnetic braking (MB) \citep{knigge_evolution_2011}.   
Eventually, MB reduces \citep{barraza-jorquera_modest_2026} or ceases entirely due to the donor star becoming fully convective \citep{1983ApJ...275..713R, 1983A&A...124..267S}, decreasing angular momentum loss from the system. This allows the bloated donor star to shrink back to thermal equilibrium and lose contact with it's Roche lobe, ending mass transfer \citep{Knigge2011, zorotovic_cataclysmic_2020, knigge_evolution_2011}. 
This is the explanation for the so-called period gap, a dearth of EM-observed systems over $f_{\rm GW}\approx$ $0.18-0.26$ mHz ($2-3$ hr orbital period). We note, however, that magnetic CVs, in which the WD has a strong magnetic field, do not have as stark of a gap \citep{schreiber_cataclysmic_2024}. 

In the period gap, the orbit shrinks due to the emission of GWs until the secondary eventually again makes contact with its Roche lobe, restarting mass transfer onto the WD. 
Once the donor's mass loss surpasses its thermal timescale to a certain degree, the orbital period will reach its minimum (maximum GW frequency). As mass loss continues, the orbital period will eventually begin to grow again due to the increasing difference between the aforementioned timescales \citep{knigge_evolution_2011, tauris_physics_2022}.
Eventually the donor star loses enough mass to become degenerate and in response to mass loss, which occurs at much lower rates than earlier stages of a CV, the donor star's radius increases.
CVs in this stage, that have evolved past the orbital period minimum with a degenerate secondary, are known as ``period bouncers'' \citep{knigge_evolution_2011}.

Previous work by \citet{Scaringi2023} explored the GW signal of the local population of CVs, finding that they will produce a unique spectral feature in the Galactic foreground, distinct from the spectrum sourced by DWDs. This local population, however, was limited to CVs within a kiloparsec (kpc) of the Solar System and did not include any systems within the period gap, which will radiate GWs as they evolve. In addition, they did not examine the detectability of the CV GW feature against a DWD foreground. We aim to understand how a Galactic CV GW signal in the LISA datastream will affect existing DWD models if neglected, and our current ability to jointly model both spectra. This will determine the breadth of CV studies possible with GWs and inform future LISA data analysis strategies to precisely characterize the foreground sourced by both CVs and other GB populations.

In \autoref{sec:methods} we discuss our synthesized CV and DWD populations as well as our GW data analysis techniques. We present the results of our analyses in \autoref{sec:results} and discussion of their implications in \autoref{sec:discussion}. We finally summarize this work in \autoref{sec:conclusion}.

\section{Methods}
\label{sec:methods}

In this work, we augment the 1-kpc local population synthesized in \citet{Scaringi2023} and generate a population representative of the MW. The GWs from these populations are injected into LISA data using the Bayesian LISA Inference Package (BLIP) \citep{banagiri_mapping_2021,  2025arXiv250820308C} which we also use for parameter estimation of the aforementioned signals. In \autoref{subsec:simulatedpops} we detail our procedure for generating CV populations, we specify the DWD datasets we use in \autoref{subsec:dwdpops}, and \autoref{subsec:gwmethods} enumerates our data analysis pipelines. To accommodate the different communities we provide the gravitational wave frequency, $f_{\rm GW}$, and the orbital period, $\rm{P}_{\rm orb}$, whenever possible, following $f_{\rm GW} = 2 f_{\rm orb} = \frac{2}{P_{\rm orb}}$.

\subsection{Simulated CV Populations}
\label{subsec:simulatedpops}
\subsubsection{Local CVs}
\label{subsubsec:1kpcsolarneighborhood}
Our local population is derived from the one presented in \citet{Scaringi2023}. Since our galaxy-wide simulation is based on their methods, we summarize the salient aspects of generating the population here.
\citet{Pala2020} curated a volume-limited sample of CVs based on parallaxes from Gaia DR2 \citep{collaboration2018gaia, collaboration2019gaia}, consisting of 42 CVs within 150 pc of the solar system. They derived a space density $\rho_{\rm CV} = 4.8^{+0.6}_{-0.8} \times 10^{-6} \rm{pc}^{-3}$.
With this space density, \citet{Scaringi2023} generate positions according to a uniform 1 kpc disk with an axisymmetric profile following:
\begin{equation}
    \rho = \rho_{0} \exp \left(-\frac{|z|}{h}\right),
\end{equation}
where $z$ is the distance above the Galactic plane and $h$ is the scale height of the CV population. While the scale height of the CV population is a function of the system age \cite{Pretorius2007}, we use the average over different CV types ($h=280$ pc) consistent with \citet{ Pala2020} and \citet{Scaringi2023}. To assign orbital periods consistent with but not limited to the observed periods of the 42 CVs, \citet{Scaringi2023} approximate the observed CDF. This approximate CDF includes the prominent features of the CV population, namely, the orbital period minimum $P_{\rm orb, min} \approx 80$~min ($f_{\rm gw} \approx 0.44$~mHz) and the period gap at $2$ hr $\lesssim \rm{P_{\rm orb}} \lesssim 3$~hr. This approximate CDF is
\begin{widetext}
\begin{equation}
    \mathrm{CDF}(x) = \begin{cases}
      0 & P_{\rm{orb}} < 76.78~\mathrm{min}, \\
      1.98 \left(\frac{x_1 + x_1^{3/4}}{1 + x_1 + x_1^2}\right) & 76.78~\mathrm{min} \leq P_{\rm{orb}} \leq 2.10~\mathrm{hr},\\
      0.834 & 2.10~\mathrm{hr} < P_{\rm{orb}} < 3.09~\mathrm{hr},\\
      0.26 \left(\frac{x_2 + x_2^{3/4}}{1 + x_2 + x_2^2}\right) + 0.834  & 3.09~\mathrm{hr} \leq P_{\rm{orb}} \leq 10.58~\mathrm{hr},\\
      1 & P_{\rm{orb}} > 10.58~\mathrm{hr}
\end{cases}.
\end{equation}
\end{widetext}
Inclinations are assigned to the systems following a uniform distribution: $\cos\iota \in U[-1, 1]$. As in \citet{Scaringi2023}, our WD mass is $m_1 \in \mathcal{N}[0.7, 0.001]$~$M_{\odot}$ and the secondary mass, $m_2$, is determined based on the evolutionary track in \citet{Knigge2011}. To match the \citet{Pala2020} sample, systems within 150 pc of the Sun are randomly picked to have their parameters substituted with the measured system parameters from the \citet{Pala2020} catalog. 

\subsubsection{Filling the EM Gap}
\label{subsubsec:fillemgap}
We augment the prescriptions in \citet{Scaringi2023} to add another layer of complexity to the populations in this work. Since the period gap is likely an observational effect, we generate detached systems that are evolving solely due to gravitational radiation. GWs will extract orbital angular momentum from the system, increasing the frequency according to \citep{1963PhRv..131..435P},

\begin{equation}
\label{eqn:gwfreqevol}
    \frac{df_{\rm GW}}{dt} = \frac{96}{5} \pi^{8/3} \frac{G^{5/3}}{c^5} \mathcal{M}^{5/3}f_{\rm GW}^{11/3}.
\end{equation}
In \autoref{eqn:gwfreqevol}, $G$ is the gravitational constant, $c$ is the speed of light, and 
\begin{equation}
    \mathcal{M} = \frac{(m_1m_2)^{3/5}}{(m_1+m_2)^{1/5}},
\end{equation}
is the chirp mass. \autoref{eqn:gwfreqevol} is for circular binaries which is assumed for all systems in the populations synthesized for this work. 

\begin{widetext}
    \begin{figure}
    \centering
    \includegraphics[width=\textwidth]{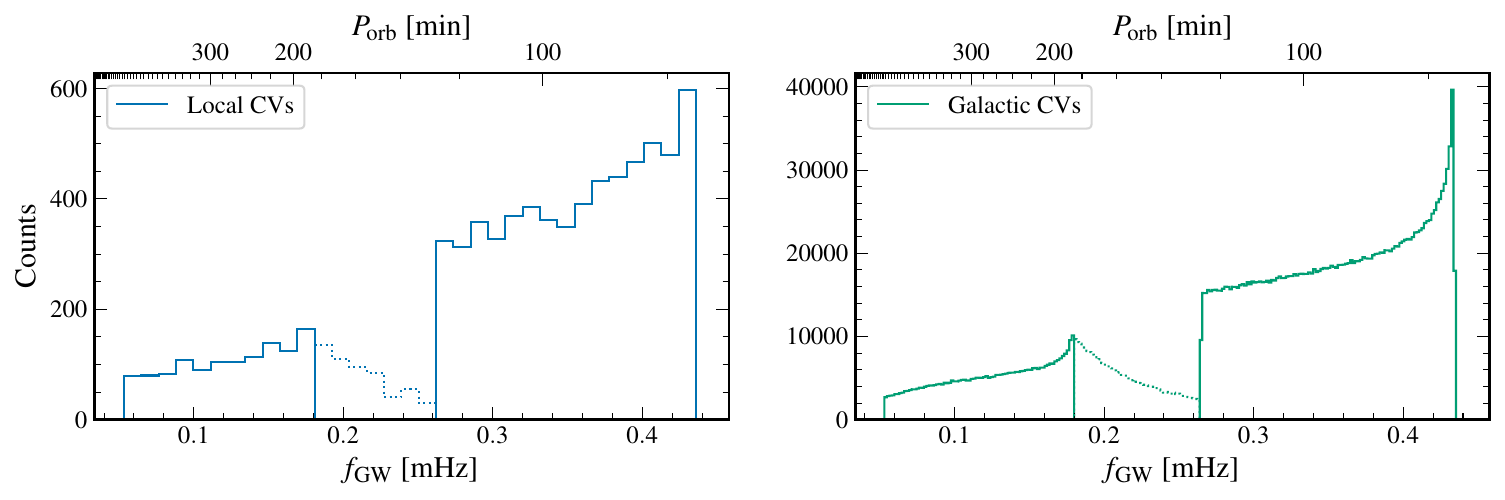}
    \caption{We show the distribution of CV GW frequency (orbital period on top axis) for both the local (left) and Galactic (right) populations. The dashed region of the histograms denote the CV period gap wherein accretion ceases, and few CVs are observed. }
    \label{fig:popdist}
\end{figure}
\end{widetext}

We synthesize systems in the period gap by first binning up all systems according to GW frequency. We determine the number of bins by using the minimum across two histogram bin estimators: Freedman-Diaconis \citep{freedman1981histogram} and Sturges rules \cite{sturges1926choice}. To avoid edge effects, we assert that there is a monotonic increase in the number of CVs in the last two bins before the EM gap. If this is not the case, we remove one bin from the previously determined number of bins. The populations' frequency (orbital period) distribution can be seen in \autoref{fig:popdist}. 
We then select systems that fall in the last frequency bin, $f'$, before the period gap and fit a kernel density estimator (KDE) to their chirp mass distribution. The KDE uses a Gaussian kernel and we utilize five-fold cross validation to select the optimal bandwidth for the kernel. Evaluating this KDE at 100 points in the range of CV chirp masses in $f'$ provides one axis of the joint distribution space from which we draw our systems.

To construct the second axis, we evaluate \autoref{eqn:gwfreqevol} at $f'$ and every frequency bin value within the period gap, over the chirp mass space. To further smooth the distribution, we perform an interpolation over the frequency axis, such that we have one hundred grid points along each axis. We perform rejection sampling in the normalized joint PDF until we have one million frequency-chirp mass pairs. To normalize the number of systems in the gap, we downsample the number of points based on the ratio of points that fall in $f'$ compared to the number from the previously described population synthesis prescription. We then only add systems to the population with frequencies above $f'+\Delta f$ where $\Delta f$ is the size of a frequency bin.

\subsubsection{The Milky Way}
\label{subsubsec:MWpop}
To place the CV population on the same footing as other existing GB datasets, we also synthesized a full Galactic population of CVs. 
We construct a simplistic galaxy model consisting of the disc and the bulge to draw galactic positions via rejection sampling. Our galactic model prescriptions are described below. The drawn positions are then normalized to match the local space density of CVs reported by \citet{Pala2020}, assumed to be constant in a 1-kpc sphere around the Solar System.
With this set of positions, we perform the same procedures as used in \autoref{subsubsec:1kpcsolarneighborhood} to draw each system's orbital period, masses, and inclinations. Similar to the procedure for the local population, we select systems within 1 kpc of the Sun to be replaced with those from the local population. This includes the 42 systems from \citet{Pala2020}. 
After creating a galaxy of CVs, we follow the procedure laid out in \autoref{subsubsec:fillemgap} to fill the EM gap in the population.

Following the models in \citet{McMillan2011} and \citet{2020ApJ...901....4B}, we assume that the MW bulge has the axisymmetric density profile
\begin{equation}
\label{eqn:bulgemod}
    \rho_{\rm b}(r') = \frac{\rho_{\rm b, 0}}{(1+r'/r_0)^{\alpha}} \exp \left[ -(r'/r_{\rm cut})^2\right],
\end{equation}
wherein, in a cylindrical coordinate system,  $r' = \sqrt{R^2+(z/q)^2}$, $r_0 = 0.075$ kpc, $\alpha = 1.8$, $r_{\rm cut} = 2.1$ kpc, and $q$, the axial ratio, is 0.5. 
Our disc component follows
\begin{equation}
\label{eqn:discmod}
    \rho_{\rm d}(R, z) = \frac{\Sigma_{\rm d, 0}}{2z_{\rm d}}\exp \left( - \frac{|z|}{z_{\rm d}}  - \frac{R}{R_{\rm d}} \right),
\end{equation}
where the scale height is $z_{\rm d} = 300$ pc for a thin disc. In our MW model the radial scale distance is $R_{\rm d} = 2.9$ kpc \citep{2020ApJ...901....4B}. 
In \autoref{eqn:bulgemod} and \autoref{eqn:discmod}, $\rho_{\rm b, 0}$ and $\Sigma_{\rm d, 0}$ are the scale and central surface density respectively. Since these are not known for CVs, they are set to unity to assemble the galactic position PDF before rejection sampling. As previously stated, we normalize the population through the local (1 kpc) CV space density, extrapolated from the 150 pc measurements of \citet{Pala2020}.

\subsection{Double White Dwarf Populations}
\label{subsec:dwdpops}
We use two DWD populations to create scenarios that are representative of existing uncertainties regarding the MW's DWD demographics. The first is that of \citet{2021MNRAS.500.4958W}, henceforth referred to as W21. We briefly summarize the details of their simulation here, but comprehensive information can be found in the original work as well as \citet{2019MNRAS.483.5518K} from which some of the simulation is based. \citet{2021MNRAS.500.4958W} use the binary population synthesis code \texttt{SeBa} developed by \citet{zwart1996population}. \citet{2001A&A...365..491N, 2001A&A...368..939N}, and \citet{toonen2012supernova} further developed DWD capabilities into this software. Parameters for the stellar population that produce the DWDs are drawn from the following distributions. The initial mass function of \citet{2001MNRAS.322..231K} is used for the primary star's mass, selecting only the range between 0.95-10 $M_{\odot}$. A uniform mass ratio distribution between zero and one is used to determine the secondary star's mass \citep{2013ARA&A..51..269D}. Finally, the population's metallicity was equivalent to the Sun with a binary fraction of 0.5. 

The \texttt{SeBa} code evolves all the initial binaries from the zero-age main sequence until DWD formation whereafter GW-driven evolution takes over \citep{1964PhRv..136.1224P}. Importantly, mass transferring binaries are removed from the population, so this dataset is entirely detached DWDs. \citet{2021MNRAS.500.4958W} assign an age distribution to the collection of binaries according to a star formation rate grid from the galaxy evolution model of \citet{1999MNRAS.307..857B}. The prescription for the binaries' common envelope parametrization is system dependent, using either the energy conservation equation or the system's angular momentum balance equation. DWD galactic positions are assigned based on three-dimensional Gaussians centered on particles within an N-body simulation that mimics the Milky Way's observed properties, originally presented in \citet{2020ApJ...890..117D}.

The second DWD population we use is derived from the fiducial FZ dataset of \citet{thiele23}, henceforth known as T23. We again briefly describe their approach for DWD population synthesis, but refer the reader to the original work for further detail. The \texttt{COSMIC} code for binary population synthesis \citep{katie_breivik_2019_3482915, cosmic2020} is used to evolve an initial stellar population, forming a catalog of detached DWD binaries. 
\citet{thiele23} implement an empirically metallicity-dependent binary fraction for solar-type stars. This relationship means that lower metallicity populations trend toward short-period binaries ($P_{\rm orb} < 10^4$ days = $f_{\rm GW} > 2$ nHz) and vice versa. Systems are assigned orbital periods according to a skewed log-normal distribution to match those found in \citet{2021MNRAS.507.3593M}. The primary mass distribution is also based on \citet{2001MNRAS.322..231K}, with mass ratios and eccentricities following a uniform distribution \citep{1992ApJ...401..265M, 1994A&A...282..801G, 2019ApJ...872..165G}. They parametrize the common envelope prescription via energy conservation and assert that orbital energy is perfectly transferred into the common envelope. 

The \citet{thiele23} population also consists of four different DWD types, which influences the population's mass and orbital period distribution. These four varieties are: two helium WDs, a carbon-oxygen WD paired with a helium WD, a double carbon-oxygen WD binary, and an oxygen-neon WD bound to either a helium, carbon-oxygen, or oxygen-neon WD. Each of the four DWD subtypes have metallicities distributed uniformly over the space $\log_{10}(\mathcal{Z}) = \rm{U}[-4, -1.52]$, where $\mathcal{Z}$ is metallicity. In this synthesized population, DWD systems are actively forming within the LISA band. This creates a spectrum that at low frequencies does not evolve due to purely GW evolution, which will necessitate bespoke spectral models for accurate characterization. We assign galactic positions according to the MW model described in  \autoref{subsubsec:MWpop} which is the galaxy model from \citet{McMillan2011} and \citet{2020ApJ...901....4B}. Inclinations are also assigned according to a uniform distribution.

\citet{thiele23} assigned metallicity-dependent ages to the DWD systems based on the m12i galaxy within the ``Latte" suite of the FIRE-2 simulations \citep{2015MNRAS.450...53H, 2016ApJ...827L..23W, 2018MNRAS.480..800H}. Changing the positions of the systems removes any associated correlation between age (and therefore frequency) and system sky location. This choice is reasonable considering that \texttt{BLIP}'s current approach for simulating a population's stochastic GW signal assumes equivalent anisotropies at every frequency, neglecting such frequency-sky-location correlations. We discuss this further in \autoref{sec:discussion}. We note that the two DWD populations considered use different procedures for assigning ages to the DWD systems, roughly representative of uncertainties regarding the MW's star formation history. 
Thus, our CV population, which does not have ages assigned due to it being an extrapolation of a local volume-limited survey, is broadly consistent with the DWD populations considered. We acknowledge, however, that this is an opportunity for improvement, considered in \autoref{sec:discussion}.

\subsection{Parameter Estimation of Gravitational Waves}
\label{subsec:gwmethods}
\subsubsection{GW Foregrounds}
\label{subsubsec:GWBs}

We are interested in the incoherent superposition of GWs sourced by each CV system. The stochastic GWF from overlapping plane waves propagating from independent sky locations at various frequencies is \citep{husa2009michele, maggiore_gravitational_2017, banagiri_mapping_2021, PhysRevD.111.023025},
\begin{equation}
\label{eqn:planewaveexp}
    h_{kl}(t, \mathbf{x}) = \sum_A \int_{-\infty}^{\infty} df \int_{S^2} d\hat{\mathbf{n}} \,\,\tilde h_A(f, \hat{\mathbf{n}}) e^{-2 \pi i f(t-\hat{\mathbf{n}}\cdot\mathbf{x}/c)}e_{kl}^A(\hat{\mathbf{n}})
\end{equation}
where $\hat{\mathbf{n}}$ is the unit vector pointing in the direction of GW propagation, $\mathbf{x}$ is the location where $h_{kl}$ is being measured, $A\in\{+, \times\}$, and $e_{kl}^A(\hat{\mathbf{n}})$ is the polarization tensor. $\tilde h_{A}(f, \hat{\mathbf{n}})$ is the Fourier domain amplitude for each GW with frequency $f$. These amplitudes obey 
\begin{equation}
    \langle \tilde h_A (f, \hat{\mathbf{n}})\tilde h_{A'}^*(f', \hat{\mathbf{n}'})\rangle = \frac{1}{2}S_A(f, \hat{\mathbf{n}}) \delta_{A, A'} \delta(f-f')\delta^2(\hat{\mathbf{n}}, \hat{\mathbf{n} '})
\end{equation}
under the assumption that the GWF is stationary, Gaussian, and unpolarized. 
$S_A(f, \hat{\mathbf{n}})$ is the one-sided GW power spectral density (PSD) for polarization $A$, but for the rest of this work we will let 
\begin{equation}
    \frac{1}{2}S_h(f, \hat{\mathbf{n}}) = S_+(f, \hat{\mathbf{n}}) = S_{\times}(f, \hat{\mathbf{n}}).
\end{equation}
Thus, $S_h$ is the total GW PSD. The signal is not isotropic since CVs will follow the spatial footprint of the MW. We assume that the spatial and spectral components of the PSD are separable:
\begin{equation}
\label{eqn:separablepsd}
    S_h(f, \hat{\mathbf{n}}) = S_h(f)P(\hat{\mathbf{n}}),
\end{equation}
where $P(\hat{\mathbf{n}})$ is the angular distribution of the GWF and satisfies
\begin{equation}
    \int_{S^2} d^2 n \, P(\hat{\mathbf{n}}) = 1.
\end{equation}
We expand on this assumption in \autoref{sec:discussion}.
Finally, we can connnect the GWF PSD to the dimensionless GW energy density:
\begin{equation}
\label{eqn:energydensityfrompsd}
    \Omega_{\rm GW} (f) = \frac{2 \pi^2}{3 H_0^2}f^3S_h(f),
\end{equation}
where $H_0$ is the Hubble constant.

We will characterize the stochastic GW signal of CVs using the LISA detector.
The measurement of \autoref{eqn:planewaveexp} in a channel $I$ of the LISA data stream is
\begin{equation}
\label{eqn:strainchanI}
        h_{I}(t, \mathbf{x}) = \sum_A \int_{-\infty}^{\infty} df \int_{S^2} d\hat{\mathbf{n}} \,\,\tilde h_A(f, \hat{\mathbf{n}}) e^{-2 \pi i ft} F_I^A(f, \hat{\mathbf{n}}).
\end{equation}
In \autoref{eqn:strainchanI}, $F_I^A(f, \hat{\mathbf{n}})$ is the antenna pattern function of channel $I$ \citep{cornish_space_2001, PhysRevD.65.022004, banagiri_mapping_2021} 
\begin{equation}
    F_I^A(f, t, \hat{\mathbf{n}}) = D_I(f, t, \hat{\mathbf{n}}):e^A(\hat{\mathbf{n}}),
\end{equation}
defined by the Michelson detector response tensor,
\begin{multline}
        D_I(f, t, \hat{\mathbf{n}}) = \frac{1}{2}[(\mathbf{u} \otimes \mathbf{u})\mathcal{T}(f, \mathbf{u}\cdot \hat{\mathbf{n}})\\
        -(\mathbf{v} \otimes \mathbf{v})\mathcal{T}(f, \mathbf{v}\cdot \hat{\mathbf{n}})]e^{2 \pi i f \, \hat{\mathbf{n}}\cdot\mathbf{r}/c}
\end{multline}
for an interferometer constructed such that the vertex is at \textbf{r} and arms oriented along unit vectors \textbf{u} and \textbf{v}. $\mathcal{T}(f, \mathbf{u}\cdot \hat{\mathbf{n}})$ is the interferometric timing transfer function, defined as \citep{1997CQGra..14.1513S, cornish_space_2001}
\begin{multline}
        \mathcal{T}(f, \mathbf{u}\cdot \hat{\mathbf{n}}) = \frac{1}{2} \bigg[ \mathcal{S} (1-\textbf{n}\cdot\textbf{u}) \exp\left(-i \frac{f}{2f_*}(3+\textbf{n}\cdot\textbf{u}) \right) \\ 
        +\mathcal{S}(1+\textbf{n}\cdot\textbf{u})\exp\left(-i \frac{f}{2f_*}(1+\textbf{n}\cdot\textbf{u})\right) \bigg],
\end{multline}
where $\mathcal{S}(x) = \mathrm{sinc}(\frac{f}{2f_*}x)$, $f_*=c/2\pi L$, and $L=2.5\times10^9$m is the LISA arm length.

To search for a GWF in our detector, we must cross-correlate the data from two LISA channels \citep{romano2017detection} to detect them. The resultant cross-correlation matrix, $C_{IJ}(f, t)$, will include the GWF, $S_{IJ}^{\rm GW}(f, t)$, and noise, $S_{IJ}^{\rm n}(f)$, PSDs for channels $I$ and $J$:
\begin{equation}
    C_{IJ}(f, t) = S_{IJ}^{\rm GW}(f, t) + S_{IJ}^{\rm n}(f).
\end{equation}
Prescriptions for $S^n_{IJ}(f)$ are described in \autoref{subsubsec:gwdataanalysis}.
Proceeding from \autoref{eqn:strainchanI}, $S_{IJ}^{\rm GW}(f,t)$ is
\begin{equation}
    S_{IJ}^{\rm GW}(f, t) = S_h(f) \times \mathcal{R}_{IJ}(f,t)
\end{equation}
where $\mathcal{R}_{IJ}(f, t)$ is the LISA response function:
\begin{equation}
\label{eqn:LISAresponsefunc}
    \mathcal{R}_{IJ}(f, t)= \frac{1}{2}\int d^2\hat{\textbf{n}} \, P(\hat{\mathbf{n}})\left( \sum_A F^A_I(f, t, \hat{\textbf{n}})F^{A*}_J(f, t, \hat{\textbf{n}})\right).
\end{equation}

\subsubsection{GW Data Analysis}
\label{subsubsec:gwdataanalysis}
While the previous section established the physical manifestation of a GWF in LISA data, here we focus on simulating LISA data and detecting GW signals within it. We first describe how \blip simulates LISA data and then our approach for analyzing it.

The time-domain LISA datastream simulated by \blip includes noise and any injected GWF. We assume rigid-body orbits of the spacecraft in the LISA constellation
\citep{colpi_lisa_2024}. The instrumental noise in the LISA data stream is simulated according to prescriptions of the power spectrum for position and acceleration noise, respectively, given in \citep{colpi_lisa_2024}:
\begin{equation}
    \label{eqn:lisanoisepsdp}
    S_{p}(f) = N_p \left[ 1 + \left( \frac{2 \, \rm mHz}{f} \right)^4\right] \rm{Hz^{-1}}
\end{equation}
\begin{equation}
    \label{eqn:lisanoisepsda}
    S_{a}(f) = \left[ 1 + \left( \frac{0.4 \, \rm mHz}{f} \right)^2\right] \left[ 1 + \left( \frac{f}{8 \, \rm{mHz}} \right)^4\right] \frac{N_a}{(2\pi f)^4}\rm{Hz^{-1}}.
\end{equation}
For all analyses in this work, we use $N_p = 9 \times 10^{-42}$ and $N_a = 3.6 \times 10^{-49} \rm{Hz}^3$ to be consistent with noise levels contained in \citep{colpi_lisa_2024}.  

We only consider two GWF sources for LISA: Galactic DWDs and CVs. For each population we transform each system's galactocentric coordinates to a Solar System Barycentric frame and retrieve their galactic latitude and longitude to place them on the sky \citep{astropy:2013, astropy:2018, astropy:2022}. We use \texttt{LEGWORK} \citep{LEGWORK_apjs, LEGWORK_joss} to calculate the GW strain of each system according to \citet{1963PhRv..131..435P}
\begin{equation}
    h = \frac{2^{22/3} G^{5/3}}{5c^4}\frac{\mathcal{M}^{5/3}(\pi f_{\rm GW})^{2/3}}{r},
\end{equation}
where $r$ is the distance to the source. We assume all sources to be monochromatic over the mission duration. Any systems that have a signal-to-noise ratio (S/N) greater than 7 are considered resolvable and are removed from the analysis. This S/N is also calculated via \texttt{LEGWORK}, using the instrumental noise and DWD foreground specified in \citep{robson_construction_2019}. This S/N ratio is orbit averaged and includes the effects of the inclination of the binary. \texttt{LEGWORK} does not perform a noise-weighted inner product, but the calculation is equivalent to an optimal S/N. This is a simplistic approach, but discriminating between resolvable and unresolvable systems is beyond the scope of this work (see \cite{littenberg_prototype_2023,  2025PhRvD.111b4060K} for work of this sort) and is more relevant for sources at higher frequencies than CVs can attain. Unresolved systems, defined as those that pass this S/N cut, are then binned by frequency and sky location. The frequency resolution is set by the LISA mission duration: $\Delta f = 1/T_{\rm obs}$. We use a four-year mission duration, resulting in frequency resolutions of $\approx 8$ nHz. The sky location binning is set by a \texttt{HEALPix} \footnote{http://healpix.sourceforge.net} skymap \citep{Zonca2019, 2005ApJ...622..759G} with a resolution of $N_{\rm pix} = 12N_{\rm side}^2$. For all analyses in this work we use $N_{\rm side}=16$, producing a resolution of 3072 pixels with each pixel containing an area of $\approx 13$ square degrees. 

\blip convolves this injected GW spectrum, with its associated spatial dependence, with the time-dependent LISA response function (see \autoref{eqn:LISAresponsefunc}
) to simulate a time series of the GWF (see \citep{rieck_stochastic_2024, PhysRevD.111.023025} for specific data simulation details). This time series is added to the simulated instrumental noise previously described to synthesize LISA data for a given channel. All analyses in this work use $X-Y-Z$ time-delay interferometry (TDI) channels \citep{1999PhRvD..59j2003T, 2000PhRvD..63b1101T}.

To analyze the simulated LISA data, \blip performs short-time Fourier transforms on segments of the datastream of length $2 \times 10^5$s. The sampling rate of the timeseries is 0.25 Hz. We focus only on Fourier modes $10^{-4}<f<10^{-2}$ Hz. \blip then implements a likelihood derived from the multidimensional complex Gaussian likelihood presented in \citet{adams_discriminating_2010}. The \blip likelihood uses the piecewise, stationary covariance matrix $C_{IJ}(f, t)$. This covariance matrix is time and frequency dependent, with $f$ and $t$ denoting the specific frequency bin and time segment. With $\tilde d_I$ being the Fourier-domain data in channel $I$ , $\vec{\theta}$ being a vector of model parameters, and $T_{\rm seg}$ being the segmentation length of the data before Fourier transformation, the \blip likelihood is \citep{adams_discriminating_2010, banagiri_mapping_2021, 2025arXiv250820308C} 
\begin{multline}
    \mathcal{L}(d|\vec{\theta}) = \prod_{f, t} \frac{1}{ T_{\rm seg}\det(2\pi C_{IJ}(f, t|\vec{\theta}))}\\
    \times \exp \left[ -\frac{2}{T_{\rm seg}} \sum_{I,J} \tilde d^*_I(f, t)(C(f,t|\vec{\theta}))^{-1}_{IJ} \tilde d_J(f,t)  \right].    
\end{multline}

With this likelihood and our uniform priors, detailed in \autoref{tab:priortable}, we perform Bayesian inference using the No-U-Turn Sampler (NUTS) from numpyro \citep{phan2019composable, bingham2019pyro} and the dynesty \citep{2004AIPC..735..395S, 10.1214/06-BA127, 2020MNRAS.493.3132S, sergey_koposov_2025_17268284} nested sampler for parameter estimation. Our models and their parameters are described in the following sections.

\smallskip
\subsubsection{Double White Dwarf Foreground Spectral Models}

Our two spectral models for the DWDs provide a range of flexibilities as required by our current understandings of the population. Our first model is the commonly-used tanh-truncated power law (hereafter known as TPL) \citep{robson_construction_2019, 2021PhRvD.104d3019K, PhysRevD.111.023025}. The truncation is necessary to accommodate DWDs being resolvable in large numbers above 1-3 mHz, driving a precipitous drop in the DWD GW spectrum. This model is defined as, 
\begin{equation}
    \Omega_{\rm GW}(f) = \frac{1}{2}\Omega_{\rm {ref}}\left(\frac{f}{f_{\rm{ref}}} \right)^{2/3} \left( 1 + \tanh\left( \frac{f_{\rm{cut}}-f}{f_{\rm{scale}}} \right) \right),
\end{equation}
where the three free parameters in the model are $\Omega_{\rm ref}$, the amplitude of the energy density, $f_{\rm{cut}}$, the cutoff frequency, and $f_{\rm{scale}}$, the frequency cutoff scaling parameter. $f_{\rm ref}$ is the reference frequency of 1 mHz and the spectral index is fixed to 2/3, the theoretical prediction for a population of binaries undergoing inspiral according to GW emission \citep{phinney_practical_2001}.

We also created a new spectral model derived from existing MW foreground models. Owing to the unique low frequency behavior of the T23 dataset, we found that the TPL spectral model did not fit the DWD population well. As such, we added two broken power laws to the model. We initially only used a single broken power law, but the data preferred using this spectral flexibility to better smooth the transition between the power law and the truncation that begins at a few mHz. The resultant model is
\begin{widetext}
    \begin{equation}
                \Omega_{\rm GW}(f) = \frac{1}{2}\Omega_{\rm ref} \left( \frac{f}{f_{\rm ref}}\right)^{\alpha_1} \left(1+\left( \frac{f}{f_{\rm b_1}}\right)^{1/\delta_1} \right)^{(\alpha_1-\alpha_2)\delta_1} \left(1+\left( \frac{f}{f_{\rm b_2}}\right)^{1/\delta_2} \right)^{(\alpha_2-\alpha_3)\delta_2} \left( 1 + \tanh \left( \frac{f_{\rm cut}-f}{f_{\rm scale}} \right) \right) ,
    \end{equation}
\end{widetext}
where $\alpha_1$, $\alpha_2$, and $\alpha_3$ are spectral indices for different power law components, $f_{\rm b_1}$ and $f_{\rm b_2}$ are the break frequencies for the two broken power law components, and $\delta_1$ and $\delta_2$ are the smoothing parameters for the two broken power laws. This double-broken, tanh-truncated power law model will be referred to as DBTPL throughout the rest of this work. This model's flexibility can create a multi-modal structure within its likelihood surface, presenting a challenge for NUTS. For this reason, we use nested sampling for any analysis that includes the DBTPL model. For all other models, we use NUTS.

\subsubsection{One-Sided Step Function Spectral Model}
We build a spectral model for the CVs that focuses mainly on the systems at frequencies higher (lower orbital periods) than the period gap. This is the strongest part of the population's GW signal and includes the frequency cutoff that exists for CVs which could aid in their identification. While more complex models for the CV population may be possible, in this work we use a simple model to attempt separating the CV signal from the DWD foreground. It is a one-sided step function of the form
\begin{equation}
    S_h(f) =  \begin{cases}
       \left( \frac{f}{0.45 \rm{mHz}} \right)^{-7/3} A_{\rm min} &  f < f_{\rm min} \\
       \left( \frac{f}{0.45 \rm{mHz}} \right)^{-7/3} A_{\rm max} & f_{\rm min} < f < f_{\rm max}\\
        0 & f > f_{\rm max}
    \end{cases},
\end{equation}
where the $f_{\rm min}$ and $f_{\rm max}$ frequency parameters define the low and high frequency edge of the mass transferring systems that have evolved beyond the period gap, and $A_{\rm min}$ and $A_{\rm max}$ are the amplitudes of the signal before $f_{\rm min}$ and after. We use 0.45 mHz as a reference frequency which is the approximate maximum frequency the CV population reaches. This spectrum is transformed into the dimensionless energy density via \autoref{eqn:energydensityfrompsd}. We do not fit all four parameters, however, defining $f_{\rm min}$ and $A_{\rm min}$ as fixed offsets from $f_{\rm max}$ and $A_{\rm max}$ respectively. These definitions are
\begin{equation}
    \log_{10}f_{\rm min} = \log_{10} f_{\rm max} - 0.22
\end{equation}
\begin{equation}
    \log_{10} A_{\rm min} = \log_{10} A_{\rm max} - 1.
\end{equation}
We choose these definitions based on the strong evidence for the end of the period gap from EM observations \citep{tauris_physics_2022, rodriguez_cataclysmic_2025, Pala2020, schreiber_cataclysmic_2024, knigge_evolution_2011}. OSS is the acronym that will refer to this this spectral model. The priors for the parameters in this model and all other spectral models are listed in \autoref{tab:priortable}.

\subsubsection{Spatial Model}
We adopt the procedure of \citet{PhysRevD.111.023025}, using a fixed spatial template in combination with our spectral models rather than searching over parameters that describe the angular distribution of GW power on the sky. Performing such a search is currently computationally prohibitive for the styles of analysis we are pursuing \citep{PhysRevD.111.023025}. We use the two-parameter MW spatial template from \citet{PhysRevD.111.023025} based on the MW model from \citet{McMillan2011} and \citet{2020ApJ...901....4B}, also used to synthesize our MW population of CVs (see \autoref{subsubsec:MWpop}) and the Galactic positions of the T23 population. The two parameters are the radial scale distance, $R_{d}$ and scale height, $z_d$, for the disk. Our templates are constructed with the true values used to create the Galactic CV population. 
In the case of W21, we do not assign the systems positions according to this model, but instead formulate an anisotropy template using the systems' true positions on the sky, as was done in \citet{PhysRevD.111.023025}. This population map anisotropy template is an exact match to the injected anisotropy. 

Hence, our spatial models for the GW signals match the prescriptions used to generate each population.
These are strong assumptions, especially considering that we expect LISA's measurements of the confusion DWD signal to constrain the Milky Way's shape \citep{2020ApJ...901....4B}. We do expect CVs to be distributed in the Galaxy similarly to DWDs though, and we would like to understand how these similar anisotropies may incur data analysis penalties when searching for multiple GB populations.  We consider alternative approaches that could be promising avenues for future work in \autoref{sec:discussion}.

\subsubsection{Statistical Measures of GW Spectra}
\label{subsubsubsec:statmeasuresgw}
\begin{figure*}
    \centering
    \includegraphics[width=\textwidth]{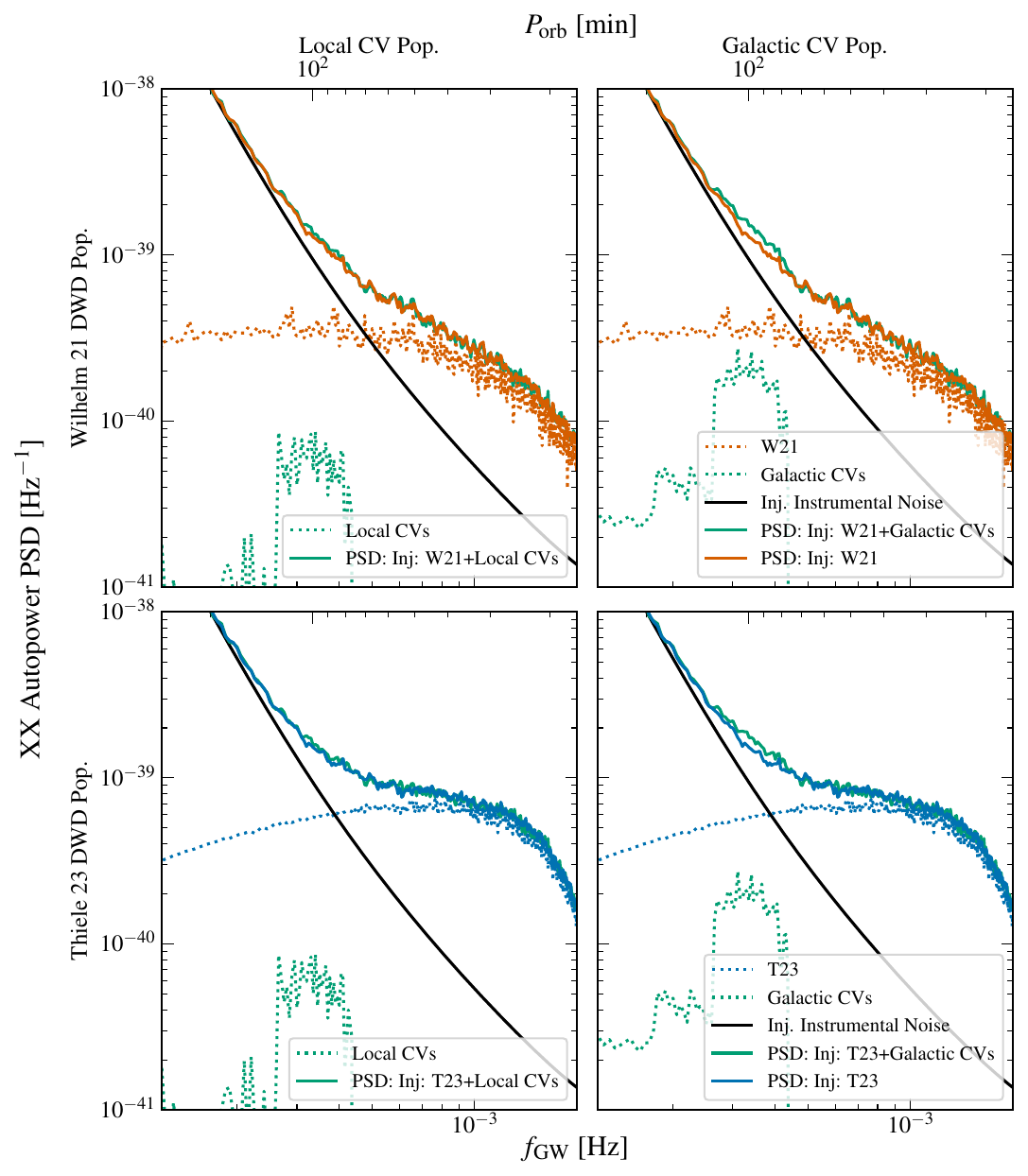}
    \caption{A comparison of the resultant PSD when injecting either the local or Galactic CV population and either the W21 or T23 DWD population into a nominal, four-year LISA mission datastream. Local (Galactic) CV population datasets are in the left (right) column, while W21 (T23) datasets are in the top (bottom) row. The green solid lines denote the recovered PSD when the CV population is present. Only the Galactic CV population creates a PSD that is substantially different from the control scenario.}
    \label{fig:fig2_psdcomp}
\end{figure*}

To determine the strength of the CV GW spectra, we calculate a single-channel, autopower S/N based on equation 10.6 from \citet{romano2017detection}:
\begin{equation}
    \rho^2 = \int_{f=0}^\infty \frac{(2fT)S^{\rm GW}_{\rm XX}(f)^2}{S_{\rm XX}^n(f)^2}d(\ln f),
\end{equation}
where $T$ is the duration of the observation and $S_{XX}^{n}(f)$ is the PSD of the noise processes. The $S^{\rm GW}_{XX}(f)$ we use to calculate the S/N of the CVs is a time-averaged one, displayed in \autoref{fig:fig2_psdcomp}. In our case, the noise PSD includes contributions from the instrument and the DWD foreground. While this equation is for isotropic stochastic GW signals, we apply it here as a conservative estimate. Incorporating the anisotropy of the signal will bring an additional boost to the S/N, as the signal strength varies throughout the datastream due to the rotation of LISA's antenna pattern.

Throughout our investigations we need to compare recovered spectra to a control scenario, wherein the CV and DWD GW spectra do not interfere with each other. To do so, we perform Bayesian inference on LISA datasets containing only W21 or T23 using either the TPL or the DBTPL spectral models. We also do this for a dataset containing the Galactic CVs and the OSS model. These recovered spectra are used as templates to measure the biases incurred when another GB population is present. 
We quantify these biases using the z-score. The z-score is defined as
\begin{equation}
    z = \left| \frac{x-\mu}{\sigma} \right|
\end{equation}
where $\mu$ is the mean and $\sigma$ is the standard deviation of the control scenario spectrum. This z-score is calculated over each frequency bin and assumes the posterior spectrum is Gaussian. It indicates the number of standard deviations a point is away from the control spectra.

\begin{table}[!ht]
\begin{ruledtabular}
\begin{tabular}{cc}
Parameter & Prior \\
\hline
 Noise&\\
 \hline
 $\log(N_a)$& U[-51, -46]\\
 $\log(N_p)$& U[-44, -39]\\

\hline
 DBTPL &\\
 \hline
$\log_{10} \Omega_{\rm ref}$& U[-22, 0]\\
$\alpha_1$& U[-4, 6]\\
$\alpha_2$& U[0, 40]\\
$\alpha_3$& U[34, 44]\\
$\delta_1$& U[0.01, 1]\\
$\delta_2$& U[0.01, 1]\\

$\log_{10} f_{\rm b_1}$& U[-3.22, -3.10]\\
$\log_{10} f_{\rm b_2}$& U[3.7, -3.22]\\

$\log_{10}(f_{\rm cut}/\rm{Hz})$& U[-3.10, -2]\\
$\log_{10}(f_{\rm scale}/\rm{Hz})$& U[-4, -2]\\

\hline
 TPL &\\
 \hline
$\log_{10} \Omega_{\rm ref}$& U[-10, -7]\\
$\log_{10}(f_{\rm cut}/\rm{Hz})$& U[-3.1, -2.4]\\
$\log_{10}(f_{\rm scale}/\rm{Hz})$& U[-4, -2]\\

\hline
OSS& \\
\hline
$\log_{10} A_{\rm max}$& U[-38, -29]\\
$\log_{10} f_{\rm max}$& U[-3.46, -3.22]\\
\end{tabular}
\end{ruledtabular}
\caption{\label{tab:priortable}PRIORS FOR MODELS USED}
\end{table}

\section{Results}
\label{sec:results}
With the simulated datasets described in \autoref{subsec:simulatedpops} and our analysis pipelines (\autoref{subsec:gwmethods}), we endeavor to understand how CVs affect DWD spectral recovery and if the two populations' GW signals can be separated from one another to extract the relevant astrophysics governing the populations. 
Our results are unambiguous: CVs can no longer be relegated to the margins of LISA science.

\subsection{The Impact of Local CVs}
\label{subsec:resultslocalcvs}
When we analyze LISA data from a nominal 4-year mission, containing our local CVs and either DWD population, we find that they make do not make a strong impact on the measured PSD. Shown in the left column of \autoref{fig:fig2_psdcomp}, the local CV population has a S/N of 45 or 41 when compared against the W21 or T23 DWD population and the instrumental noise.
We note that the strength and shape of the local CV signal differs from the results reported in \citet{Scaringi2023}. Discussion of these results in context with this previous work can be found in \autoref{sec:discussion}. The local CV population GW spectrum is also jagged due to it being dominated by shot noise from nearby, unresolvable systems, lacking a strong underlying GWF. We turn our focus to the Galactic population of CVs, since limiting our perspective to local CVs is not an authentic representation of what the LISA detector will observe.

\subsection{CVs in the Milky Way}
\label{subsec:resultsmw}
While the local CV population sources a minimal feature, our Galactic population produces a noticeable signal in LISA data, regardless of which DWD population is present. We compare the observed PSD for datasets with Galactic CVs and without them on the right side of \autoref{fig:fig2_psdcomp}. For both W21 (top) and T23 (bottom), the MW's CVs create a distinct spectral feature visible in the observed PSD, with S/N of 79 and 73 for W21 and T23 respectively.
We note that the Galactic population of CVs has a smoother spectrum, wherein the impact of nearby unresolved sources is lessened.

Part of our motivation for this study was discerning the degree to which CVs, a previously ignored class of GB, can affect current foreground spectral models. With our statistical techniques, described in \autoref{subsubsubsec:statmeasuresgw}, we can compare the spectral recovery from datasets containing a DWD population with and without the CVs present, measuring the bias incurred by neglecting CVs.
For the datasets using W21 as the DWD population, we use the TPL model, due to its GW-driven inspiral at low frequencies. The T23 dataset's unique low-frequency behavior requires a more flexible model, so we use our DBTPL spectral model for datasets which include it. These choices are precluded on knowing which model is a better fit to the true DWD spectrum, an assumption we relax in \autoref{subsec:modelingcvs}. Further discussion of these modeling choices is found in \autoref{sec:discussion}.

\begin{figure}[h]
    \centering
    \includegraphics[width=\columnwidth]{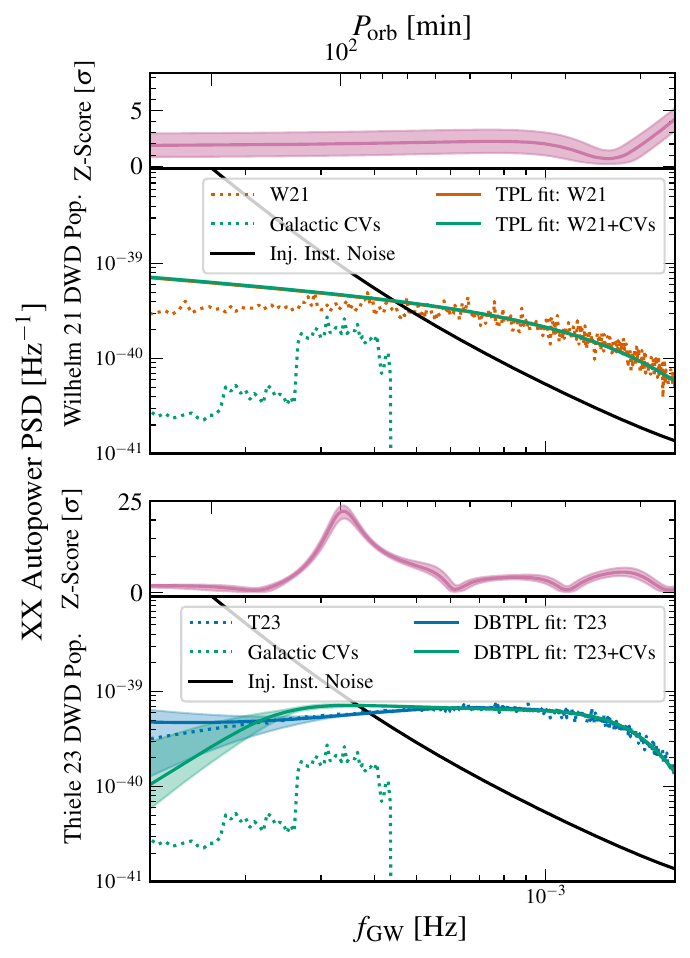}
    \caption{A comparison of the posterior spectrum recovered from a control dataset containing a DWD population (orange from W21 and blue from T23) and when the CV population is present (green). Shaded regions denote 95\% and 68\% CI for spectra and z-scores respectively. The top panels of each plot show the z-score with respect to frequency, measuring the difference between the green spectra and those recovered from a DWD-only dataset.}
    \label{fig:fig3_dwdspecrecovery}
\end{figure}

When we apply the TPL model to a dataset containing the W21 population of DWDs and the Galactic CVs, we recover a slightly biased spectrum. While the TPL is relatively rigid, with a fixed $\alpha=2/3$ spectral index power law at low frequencies, the GW power contributed by the CVs causes an overestimate of the power law amplitude. This results in a spectrum that at low-frequencies ($f_{\rm GW} < 0.5 \rm{mHz}$) is $2.0\pm1.1  \sigma$ (median and 68\% credible interval (CI)) from the spectrum recovered in a DWD-only dataset. This impact is despite the fact that the CVs are limited to a very narrow range of frequencies. A comparison of the recovered TPL spectra from datasets which do or do not include the Galactic CVs can be found in the top of \autoref{fig:fig3_dwdspecrecovery}. We also show the z-scores for the spectrum recovered from a CV and W21 dataset, relative to the recovery for a control dataset with only W21.

The Galactic CVs have a much larger impact on the DBTPL model due to its inherent flexibility. Since this model was constructed to accommodate the low frequency behavior of the T23 DWDs, the GWs from CVs drastically alter the recovered spectrum. At low frequencies, the z-score is $8.2^{+8.4}_{-6.7}\sigma$. Comparisons between the spectra recovered from a control, T23-only dataset and one that includes the Galactic CVs, as well as the z-scores can be found in the bottom of \autoref{fig:fig3_dwdspecrecovery}. 

These effects on foreground models may not be as pronounced in a realistic noise environment. When LISA flies, there are likely to be data gaps, glitches, and instrumental performance variations that could alter the low-frequency behavior of the noise curve \citep{colpi_lisa_2024}. 
The narrow frequency range over which CVs are present and their limited strength do make them vulnerable to becoming a threshold source. If their GWs are not absorbed by the DWD model in such a case, they will leave residual GW power in the datastream, potentially hindering data analysis schemes for other source classes.

\subsection{Modeling the CV Signal}
\label{subsec:modelingcvs}

\begin{figure*}
    \centering
    \includegraphics[width=\textwidth]{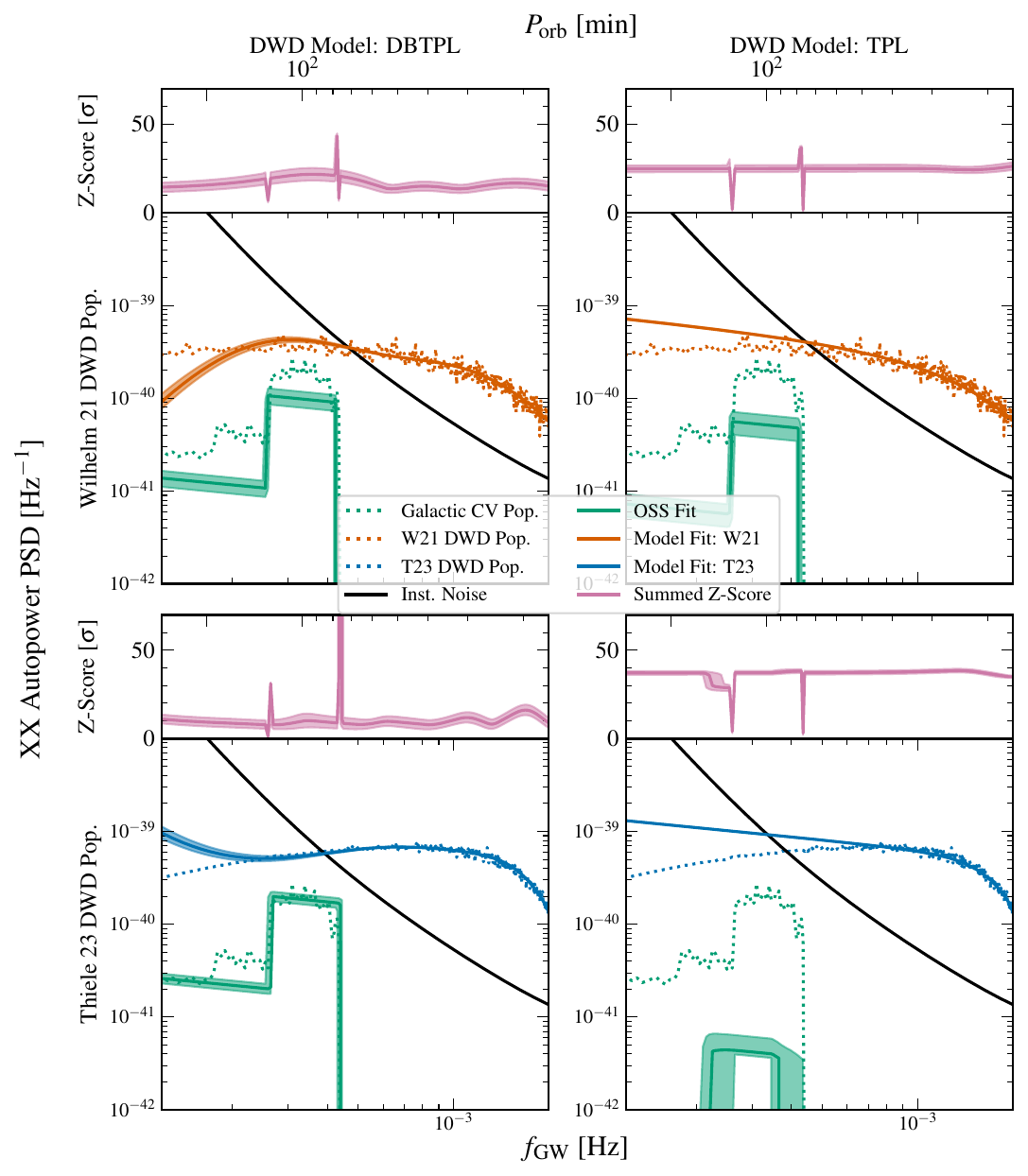}
    \caption{Shown are the posteriors from the two model prescriptions when applied to our two datasets containing CVs and DWDs. The best spectral separation occurs when DBTPL+OSS is applied to a dataset containing the T23 DWDs and CVs (bottom left). Discontinuous increases or decreases in the summed z-score correspond to where the step function in the control scenario turns on or off, creating large disagreements if the OSS frequency is misestimated.}
    \label{fig:fig4_sepcomp}
\end{figure*}

We relax our previous simplifying assumption in our modeling, wherein we know the spectral model that best fits the DWDs accompanying the CVs, in an attempt to explore the prospects for jointly modeling both populations. In this section, we consider two LISA datasets with Galactic CVs: one with the W21 DWD population and one with the population from T23. We then apply our two signal prescriptions to each dataset to understand how our inference depends on our modeling choices and explore prospects for spectral separation. Our models consist of our OSS spectral model for the Galactic CVs and either of the two DWD spectral models (TPL or DBTPL). The posterior spectra for these analyses is shown in \autoref{fig:fig4_sepcomp}. Since we are fitting both populations, we have z-scores that denote the accuracy of the OSS recovery as well as that of the DWD spectral model. To condense these results, we report the sum of these two z-scores, signifying how well the inference does overall at recovering unbiased spectra.

We find that current approaches struggle to accurately isolate the respective spectra. 
When the TPL spectral model is present (right side of \autoref{fig:fig4_sepcomp}), it overestimates the GW power contribution from DWDs, resulting in an underestimated step function amplitude. This underestimation is worst in the case of the T23 dataset where the DWD spectra falls off much more steeply than the 2/3 power law asserted in the model. Consequently, the (low-frequency) z-score for the T23 scenario is $37.2^{+1.2}_{-1.5} \sigma$ versus $24.9\pm2.1 \sigma$ for a dataset containing W21 DWDs instead.
Despite this, these prescriptions return reasonably accurate frequency estimations for the OSS model. When we apply the DBTPL+OSS prescription to the CV dataset with W21 DWDs, we find step function amplitude estimation to be more accurate than either of the cases including the TPL model.
In this circumstance, the DBTPL still seems to be absorbing some of the CV GW power and is a very poor fit to the W21 spectrum at low frequencies. The effect is a  z-score below 0.5 mHz of $18.7\pm4.5 \sigma$.

When we apply the DBTPL and OSS models to a dataset with the T23 DWDs and the CVs we are able to recover each spectra more accurately than other scenarios. In the low-frequency regime, the combined z-score for the two spectra is  $z=9.2^{+3.6}_{-3.2}\sigma$. The step function returns an accurate amplitude and frequency estimation and the DBTPL is a reasonable fit to the T23 spectrum. At very low frequencies though, the recovered DBTPL spectrum is not consistent with the T23 population. 
Biased spectral recoveries will lead to incorrect astrophysical conclusions and may also leave extraneous power in the LISA datastream to corrupt parameter estimation for other source classes. Thus, model improvements that take into account the contributions from different GB populations that source the Galactic foreground are of great importance. The precipitous increases and decreases in the z-score seen throughout the various analysis configurations are due to the discontinuous nature of the step function. These points correspond to where the step function in the control scenario turns on or off, meaning that slight step function frequency misestimations create discontinuous points.
We elaborate on this limitation in \autoref{sec:discussion}.

\section{Discussion}
\label{sec:discussion}

\subsection{Contextualizing Previous Results}
Our results for the local population of CVs do not agree with those of \citet{Scaringi2023}. We find a different spectral shape for the local CVs and do not find that they make an impact on LISA observations. One major difference between these works is that their DWD population is from the Sangria version of the LISA data challenges \citep{le_jeune_2022_7132178}. We also find that there are no resolvable ($\rm{S/N}>7$) CV systems. 
This indicates that GWs will be used almost entirely to study the GWF created by the population, rather than individual systems. This is not a definitive or exhaustive examination of the resolvability of individual CV systems though and further work to explore this subject in detail is necessary.

\begin{figure}[h!]
    \centering
    \includegraphics[width=\columnwidth]{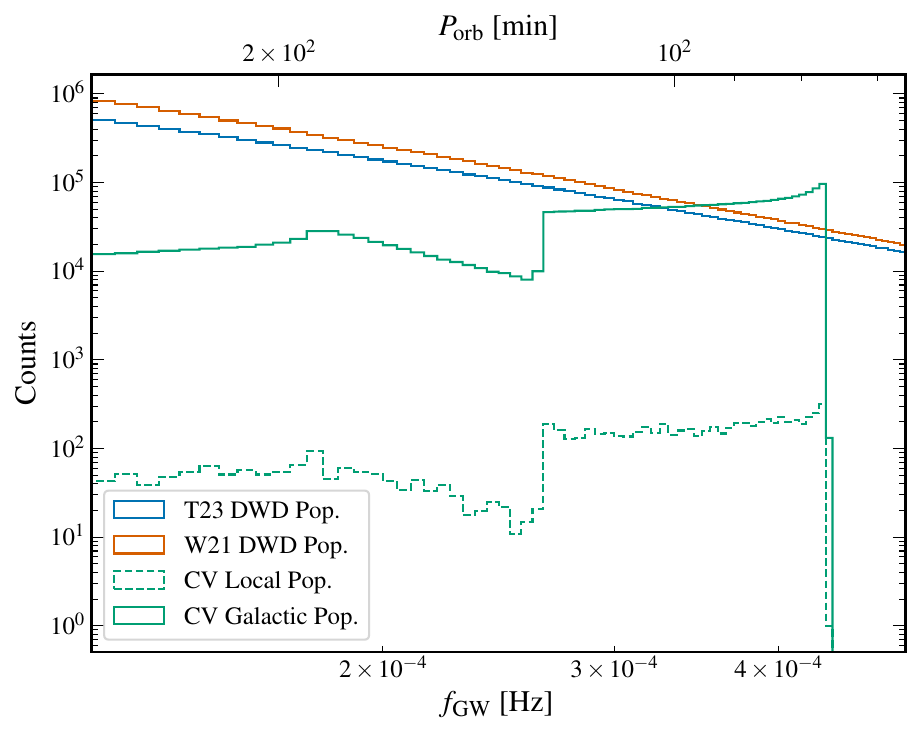}
    \caption{A histogram of the number of binaries present at each frequency for all GB populations considered in this work. When considering the whole galaxy, CVs beyond the period gap can outnumber DWDs in certain frequency bins.}
    \label{fig:occfrac}
\end{figure}

We have determined that the source of the discrepancy between our results and that of \citet{Scaringi2023} is due to a miscalculation in \citet{Scaringi2023} in which total mass was used instead of chirp mass in calculating the GW strain of each CV system. We also show the number of systems per frequency bin for the different populations examined in this work in \autoref{fig:occfrac}. This makes it easier to see why the local population makes a much smaller impact when pitted against the large number of DWDs present in the Milky Way. A Galactic population of CVs, however, has a comparable or even greater number of systems than the DWD populations considered here, depending on the frequency bin.

\subsection{Implications and Extensions}
In \autoref{subsec:resultsmw} we found that the GWF created by the Galactic population of CVs is strong and can bias models for the DWD population if ignored. We note that the high frequency portion, composed of systems that have evolved beyond the period gap, is the most impactful part of the spectrum. This is unsurprising considering that both a majority of CVs exist in this frequency range \citep{Pala2020, 2009MNRAS.397.2170G}, and LISA's instrumental noise weakens over the same region. Whether there is enough information in this limited view of the spectrum to resolve ongoing disagreements between theory and observation regarding the distribution of CVs with respect to the period gap \citep{1993A&A...271..149K, 2001ApJ...550..897H, 2015ApJ...809...80G, Pala2020}
remains to be seen. 

We found in \autoref{subsec:modelingcvs} that current techniques are unable to accurately characterize both the CV GW spectrum and that of the DWDs. 
Cases using the TPL spectral model struggle due to it being a poor fit to both DWD GW spectra. This causes the CV GW feature to be absorbed into the TPL. Model prescriptions including a DBTPL perform better, getting more accurate CV spectrum recoveries. The W21 case does exhibit some degree of DBTPL absorbing CV GW power though.  Across both DWD populations, the DBTPL recovery disagrees with the true spectrum at very low frequencies. There are a few factors that could be causing this. One element may be the spectra falling under the rising noise floor of the detector, making it more difficult to resolve them accurately. In a control scenario, however, the low frequencies of the T23 spectrum are within the recovered 95\% CI of the DBTPL. W21 is a more challenging case due to its spectrum being weaker overall than T23. 
The simplicity of the OSS spectral model could also be a limiting factor, as a spectral model derived from CV population simulations could aid recovery.
These factors point toward the inclusion of a CV feature in Galactic foreground models, to accommodate the multiple stellar binary populations that source the overall confusion signal. Including a step function within either the TPL or DBTPL could improve the recovery of both components, but such a modification could also add model degeneracies due to the increased flexibility. Incorporating EM-derived priors to prevent this will be critical for the success of such multi-component models. Further study is additionally needed to understand how to derive astrophysical meaning from these hybrid models. 
These results underscore the intertwined nature of multi-GB population inference within LISA data.

There are several opportunities for future work that may simultaneously improve prospects for spectral separation and our ability to derive astrophysical insight from the CV GW spectrum.
One of these is the incorporation of more realistic astrophysics into our CV population.
A large factor in the strength of the spectrum is the overall space density of CVs in the Galaxy. For consistency with \citet{Scaringi2023}, we use the space density derived in \citet{Pala2020} based on CVs in the closest 150 pc. This local space density is used to calibrate an overall number of CVs in the Galaxy (See \autoref{subsubsec:MWpop})
This measurement is the sole data point from which we extrapolate the number of CVs in the Galaxy, so any change (e.g. \citet{rodriguez_cataclysmic_2025} found a slightly lower CV space density) will change the number of CVs contributing to the spectrum, and therefore its strength. Our population synthesis also uses a frequency (orbital period) distribution that best matches non-magnetic CVs. Magnetic CVs, where the WD has a strong magnetic field, have a smaller or even non-existent period gap due to the WD's magnetic field weakening magnetic braking \citep{belloni_evidence_2020, schreiber_cataclysmic_2024}. Depending on the true prevalence of magnetic CVs within the overall Galactic CV population (\citet{Pala2020} measure $36\pm7$ percent of CVs are magnetic), the low frequency edge of the spectrum, at the end of the canonical period gap, may be smoother.

A suite of simulated CV populations that encompass these uncertainties and others would aid in creating astrophysically-motivated spectral templates that directly tie the GW spectrum to physical parameters. Of special interest would be whether the high frequency slope of the spectrum can provide evidence to adjudicate between proposed theories for what process is removing angular momentum in addition to GWs at high frequencies (short orbital periods) \citep{schreiber_three_2016, zorotovic_cataclysmic_2020, tauris_physics_2022, metzger_transients_2021}.
It would also be beneficial to explore consistent binary population synthesis prescriptions for both the CV population and DWDs. Understanding covariances between these two populations and physically-driven spectral models could improve the astrophysical insights gleaned from inference-within-global-fit schemes as demonstrated by \citet{criswell_foundation_2026} and \citet{2026PhRvD.113l3006D}.

The pipeline used for Bayesian inference could also be improved to aid in spectral separation. \blip does not currently have the capability to perform inference on models that contain frequency-dependent anisotropies. We believe these spatio-spectral models, i.e. discarding the assumption made in \autoref{eqn:separablepsd}, could be a promising course for future work since older CVs, with higher GW frequencies, are expected to be distributed with a larger scale height than younger, lower frequency ones \citep{pretorius_influence_2007}. Such spatio-spectral models could also be used to model the various types of DWDs according to their different expected spatial distributions \citep{lamberts_predicting_2019}. Previous work by \citet{PhysRevD.111.023025} has found that the unique anisotropies of coincident GWFs provide powerful leverage for differentiating their respective spectra.
These spatio-spectral models would strengthen this effect by exploiting the frequency dependence of each population's anisotropy to improve spectral separation. This is apt for cases like the CV and DWD populations, where both ostensibly have similar anisotropies, but contain vastly different frequency-sky-location correlations. We intend to explore the capabilities of spatio-spectral models in a future study.

\section{Conclusion}
\label{sec:conclusion}
Extant gaps in the understanding of the physics governing CVs can only be answered via population-level studies \citep{rodriguez_cataclysmic_2025}, giving GWs a robust perspective from which to glean answers. 
Previous work examined the upcoming LISA mission's GW observations of a local population of CVs \citep{Scaringi2023}. 
We expand on this previous work, motivated to answer the following questions: what might the GW spectrum of a Galactic population of CVs look like? Can one extract the GW spectrum of CVs from a foreground of DWDs? Does this depend on the DWD population or the modeling assumptions? The answers to these questions will determine the landscape of CV astrophysics made possible by data from the LISA mission. 

To answer the aforementioned questions, we first updated the local CV population from \citet{Scaringi2023} and expanded the methods to generate a Galactic population (See \autoref{subsec:simulatedpops} for details). 
We also selected two DWD populations from the literature to represent the variety of predictions for their predicted GW foreground: \citet{2021MNRAS.500.4958W} (W21) and \citet{thiele23} (T23). We then use the \blip software suite to simulate nominal, 4-year mission LISA data including the GWs from these populations. We also use \blip to then perform Bayesian inference upon these datasets. 

In contrast to \citet{Scaringi2023}, we found that the GW signal of a local population of CVs cannot be detected against either of the DWD foregrounds considered in this work. A Galactic CV population, however, is much stronger, leaving a unique imprint on the recovered PSD (See \autoref{fig:fig2_psdcomp} for comparison.) If ignored, it will bias DWD spectral models, seen in \autoref{fig:fig3_dwdspecrecovery}.

Astrophysical meaning can only be derived if the CV GW spectrum can be recovered, however, so we built a simple spectral model to fit to the observed spectrum (CV and DWD spectral models described in \autoref{subsec:gwmethods}). We then fit datasets containing the Galactic CVs and a DWD population to characterize our ability to separate their spectra. While some signal prescriptions perform better than others and can recover the CV GW spectrum, none recover both spectra accurately. 
These difficulties are due to DWD models absorbing the CV GW feature or being a poor match to the DWD population present, preventing the CV spectral model from accurately characterizing the signal.
Exacerbating the problem, both populations possess equivalent anisotropies, preventing a modeling approach that takes advantage of such a difference to separate their spectra \citep{PhysRevD.111.023025}.

These results indicate a challenging yet rewarding path for the future of CV and Galactic foreground GW astrophysics. 
Similar to other source classes in LISA, CV science is inherently intertwined with that of DWDs. 
The creation of multi-component spectral models encompassing the contributions of multiple GB populations, including CVs, will be critical for accurate characterization of the Galactic foreground. 
Further study is also needed to connect the GW measurements of CVs to the processes dictating their evolution. These opportunities, and others noted in \autoref{sec:discussion}, will prepare the community for the release of true LISA data, enabling novel astrophysical insight into the complex lives of cataclysmic variables.

\section*{Acknowledgments}
The authors would like to thank Valeriya Korol and Shane Larson for their insightful suggestions which improved the quality of this article. We further appreciate Valeriya Korol for her support regarding her DWD dataset which did not end up in the final version of this work. We also express gratitude toward Siqi Zhang and Simone Scaringi for finding a mistake in our software prior to the submission of this manuscript. This work began in earnest at the 2025 LISA Sprint in Huntsville, AL, so we also thank the organizers of the workshop. 
LS would like to thank Kyle Gersbach, Manuel Pichardo Marcano, William Lamb, and Matt Miles for their ideas and helpful discussions. LS is greatly appreciative of Annie Parnell for moral support and finding opportunities for enjoyable activities outside of astrophysics throughout the preparation of this manuscript. LS is greatly appreciative of support from NSF AST2307719 and NRT-2125764. AWC acknowledges support from NSF NRT-2125764 and NASA LPS-80NSSC26K0342. SRT acknowledges support from an NSF CAREER \#2146016, NSF AST-2307719, NSF NRT-2125764, and NASA LPS-80NSSC26K0342. SRT also acknowledges support from a Chancellor's Faculty Fellowship from Vanderbilt University. 
KB acknowledges support from the Falco DeBenedetti Career Development Professorship. 
This work was conducted in part using the resources of the Advanced Computing Center for Research and Education (ACCRE) at Vanderbilt University, Nashville, TN. This work utilized the software suites \texttt{astropy} \cite{astropy:2013, astropy:2018, astropy:2022}, \texttt{corner} \cite{corner}, \texttt{HEALPix} \cite{2005ApJ...622..759G}, \texttt{healpy} \cite{Zonca2019}, \texttt{Jupyter} \cite{soton403913}, \texttt{Matplotlib} \cite{Hunter:2007}, \texttt{Numpy} \cite{harris2020array}, \texttt{JAX} \citep{jax2018github},  and \texttt{Scipy} \cite{2020SciPy-NMeth}.

\nocite{*}

\bibliography{apssamp}

@ARTICLE{Scaringi2023,
       author = {{Scaringi}, S. and {Breivik}, K. and {Littenberg}, T.~B. and {Knigge}, C. and {Groot}, P.~J. and {Veresvarska}, M.},
        title = "{Cataclysmic variables are a key population of gravitational wave sources for LISA}",
      journal = {mnras},
         year = 2023,
        month = oct,
       volume = {525},
       number = {1},
        pages = {L50-L55},
          doi = {10.1093/mnrasl/slad093},
archivePrefix = {arXiv},
       eprint = {2307.02553},
 primaryClass = {astro-ph.HE},
       adsurl = {https://ui.adsabs.harvard.edu/abs/2023MNRAS.525L..50S}
}

@ARTICLE{Pala2020,
       author = {{Pala}, A.~F. and {G{\"a}nsicke}, B.~T. and {Breedt}, E. and {Knigge}, C. and {Hermes}, J.~J. and {Gentile Fusillo}, N.~P. and {Hollands}, M.~A. and {Naylor}, T. and {Pelisoli}, I. and {Schreiber}, M.~R. and {Toonen}, S. and {Aungwerojwit}, A. and {Cukanovaite}, E. and {Dennihy}, E. and {Manser}, C.~J. and {Pretorius}, M.~L. and {Scaringi}, S. and {Toloza}, O.},
        title = "{A Volume-limited Sample of Cataclysmic Variables from Gaia DR2: Space Density and Population Properties}",
      journal = {mnras},
         year = 2020,
        month = may,
       volume = {494},
       number = {3},
        pages = {3799-3827},
          doi = {10.1093/mnras/staa764},
archivePrefix = {arXiv},
       eprint = {1907.13152},
 primaryClass = {astro-ph.SR},
       adsurl = {https://ui.adsabs.harvard.edu/abs/2020MNRAS.494.3799P}
}

@ARTICLE{Pretorius2007,
       author = {{Pretorius}, M.~L. and {Knigge}, C. and {O'Donoghue}, D. and {Henry}, J.~P. and {Gioia}, I.~M. and {Mullis}, C.~R.},
        title = "{The space density of cataclysmic variables: constraints from the ROSAT North Ecliptic Pole survey}",
      journal = {mnras},
         year = 2007,
        month = dec,
       volume = {382},
       number = {3},
        pages = {1279-1288},
          doi = {10.1111/j.1365-2966.2007.12461.x},
archivePrefix = {arXiv},
       eprint = {0709.1887},
 primaryClass = {astro-ph},
       adsurl = {https://ui.adsabs.harvard.edu/abs/2007MNRAS.382.1279P}
}

@ARTICLE{1963PhRv..131..435P,
       author = {{Peters}, P.~C. and {Mathews}, J.},
        title = "{Gravitational Radiation from Point Masses in a Keplerian Orbit}",
      journal = {Physical Review},
         year = 1963,
        month = jul,
       volume = {131},
       number = {1},
        pages = {435-440},
          doi = {10.1103/PhysRev.131.435},
       adsurl = {https://ui.adsabs.harvard.edu/abs/1963PhRv..131..435P}
}

@ARTICLE{Taylor2021,
       author = {{Taylor}, Stephen R.},
        title = "{The Nanohertz Gravitational Wave Astronomer}",
      journal = {arXiv e-prints},
         year = 2021,
        month = may,
          eid = {arXiv:2105.13270},
        pages = {arXiv:2105.13270},
          doi = {10.48550/arXiv.2105.13270},
archivePrefix = {arXiv},
       eprint = {2105.13270},
 primaryClass = {astro-ph.HE},
       adsurl = {https://ui.adsabs.harvard.edu/abs/2021arXiv210513270T}
}

@ARTICLE{2025arXiv250820308C,
       author = {{Criswell}, Alexander W. and {Banagiri}, Sharan and {Lawrence}, Jessica and {Schult}, Levi and {Rieck}, Steven and {Taylor}, Stephen R. and {Mandic}, Vuk},
        title = "{Flexible Spectral Separation of Multiple Isotropic and Anisotropic Stochastic Gravitational Wave Backgrounds in LISA}",
      journal = {arXiv e-prints},
         year = 2025,
        month = aug,
          eid = {arXiv:2508.20308},
        pages = {arXiv:2508.20308},
          doi = {10.48550/arXiv.2508.20308},
archivePrefix = {arXiv},
       eprint = {2508.20308},
 primaryClass = {astro-ph.IM},
       adsurl = {https://ui.adsabs.harvard.edu/abs/2025arXiv250820308C}
}

@ARTICLE{Knigge2011,
       author = {{Knigge}, Christian and {Baraffe}, Isabelle and {Patterson}, Joseph},
        title = "{The Evolution of Cataclysmic Variables as Revealed by Their Donor Stars}",
      journal = {apjs},
         year = 2011,
        month = jun,
       volume = {194},
       number = {2},
          eid = {28},
        pages = {28},
          doi = {10.1088/0067-0049/194/2/28},
archivePrefix = {arXiv},
       eprint = {1102.2440},
 primaryClass = {astro-ph.SR},
       adsurl = {https://ui.adsabs.harvard.edu/abs/2011ApJS..194...28K}
}

@article{freedman1981histogram,
  title={On the histogram as a density estimator: L 2 theory},
  author={Freedman, David and Diaconis, Persi},
  journal={Zeitschrift f{\"u}r Wahrscheinlichkeitstheorie und verwandte Gebiete},
  volume={57},
  number={4},
  pages={453--476},
  year={1981},
  publisher={Springer-Verlag Berlin/Heidelberg}
}

@article{sturges1926choice,
  title={The choice of a class interval},
  author={Sturges, Herbert A},
  journal={Journal of the american statistical association},
  volume={21},
  number={153},
  pages={65--66},
  year={1926},
  publisher={New York}
}

@ARTICLE{McMillan2011,
       author = {{McMillan}, Paul J.},
        title = "{Mass models of the Milky Way}",
      journal = {mnras},
         year = 2011,
        month = jul,
       volume = {414},
       number = {3},
        pages = {2446-2457},
          doi = {10.1111/j.1365-2966.2011.18564.x},
archivePrefix = {arXiv},
       eprint = {1102.4340},
 primaryClass = {astro-ph.GA},
       adsurl = {https://ui.adsabs.harvard.edu/abs/2011MNRAS.414.2446M}
}

@ARTICLE{2020ApJ...901....4B,
       author = {{Breivik}, Katelyn and {Mingarelli}, Chiara M.~F. and {Larson}, Shane L.},
        title = "{Constraining Galactic Structure with the LISA White Dwarf Foreground}",
      journal = {\apj},
         year = 2020,
        month = sep,
       volume = {901},
       number = {1},
          eid = {4},
        pages = {4},
          doi = {10.3847/1538-4357/abab99},
archivePrefix = {arXiv},
       eprint = {1912.02200},
 primaryClass = {astro-ph.GA},
       adsurl = {https://ui.adsabs.harvard.edu/abs/2020ApJ...901....4B}
}

@ARTICLE{LEGWORK_apjs,
    author = {{Wagg}, T. and {Breivik}, K. and {de Mink}, S.~E.},
    title = "{LEGWORK: A Python Package for Computing the Evolution and Detectability of Stellar-origin Gravitational-wave Sources with Space-based Detectors}",
    journal = {apjs},
    year = 2022,
    month = jun,
    volume = {260},
    number = {2},
    eid = {52},
    pages = {52},
    doi = {10.3847/1538-4365/ac5c52},
    archivePrefix = {arXiv},
    eprint = {2111.08717},
    primaryClass = {astro-ph.HE},
    adsurl = {https://ui.adsabs.harvard.edu/abs/2022ApJS..260...52W}
}

@ARTICLE{LEGWORK_joss,
    author = {{Wagg}, Tom and {Breivik}, Katelyn and {de Mink}, Selma},
    title = "{LEGWORK: A python package for computing the evolution and detectability of stellar-origin gravitational-wave sources with space-based detectors}",
    journal = {The Journal of Open Source Software},
    year = 2022,
    month = feb,
    volume = {7},
    number = {70},
    eid = {3998},
    pages = {3998},
    doi = {10.21105/joss.03998},
    adsurl = {https://ui.adsabs.harvard.edu/abs/2022JOSS....7.3998W}
}

@article{husa2009michele,
  title={Michele Maggiore: Gravitational waves. Volume 1: theory and experiments},
  author={Husa, Sascha},
  journal={General Relativity and Gravitation},
  volume={41},
  number={7},
  pages={1667--1669},
  year={2009},
  publisher={Springer Netherlands}
}

@ARTICLE{2014LRR....17....3P,
       author = {{Postnov}, Konstantin A. and {Yungelson}, Lev R.},
        title = "{The Evolution of Compact Binary Star Systems}",
      journal = {Living Reviews in Relativity},
         year = 2014,
        month = dec,
       volume = {17},
       number = {1},
          eid = {3},
        pages = {3},
          doi = {10.12942/lrr-2014-3},
archivePrefix = {arXiv},
       eprint = {1403.4754},
 primaryClass = {astro-ph.HE},
       adsurl = {https://ui.adsabs.harvard.edu/abs/2014LRR....17....3P}
}

@article{amaro2017laser,
  title={Laser interferometer space antenna},
  author={Amaro-Seoane, Pau and Audley, Heather and Babak, Stanislav and Baker, John and Barausse, Enrico and Bender, Peter and Berti, Emanuele and Binetruy, Pierre and Born, Michael and Bortoluzzi, Daniele and others},
  journal={arXiv preprint arXiv:1702.00786},
  year={2017}
}

@ARTICLE{thiele23,
       author = {{Thiele}, Sarah and {Breivik}, Katelyn and {Sanderson}, Robyn E. and {Luger}, Rodrigo},
        title = "{Applying the Metallicity-dependent Binary Fraction to Double White Dwarf Formation: Implications for LISA}",
      journal = {\apj},
         year = 2023,
        month = mar,
       volume = {945},
       number = {2},
          eid = {162},
        pages = {162},
          doi = {10.3847/1538-4357/aca7be},
archivePrefix = {arXiv},
       eprint = {2111.13700},
 primaryClass = {astro-ph.HE},
       adsurl = {https://ui.adsabs.harvard.edu/abs/2023ApJ...945..162T}
}

@ARTICLE{cosmic2020,
       author = {{Breivik}, Katelyn and {Coughlin}, Scott and {Zevin}, Michael and {Rodriguez}, Carl L. and {Kremer}, Kyle and {Ye}, Claire S. and {Andrews}, Jeff J. and {Kurkowski}, Michael and {Digman}, Matthew C. and {Larson}, Shane L. and {Rasio}, Frederic A.},
        title = "{COSMIC Variance in Binary Population Synthesis}",
      journal = {\apj},
         year = 2020,
        month = jul,
       volume = {898},
       number = {1},
          eid = {71},
        pages = {71},
          doi = {10.3847/1538-4357/ab9d85},
archivePrefix = {arXiv},
       eprint = {1911.00903},
 primaryClass = {astro-ph.HE},
       adsurl = {https://ui.adsabs.harvard.edu/abs/2020ApJ...898...71B}
}

@article{PhysRevD.111.023025,
  title = {Templated anisotropic analyses of the LISA Galactic foreground},
  author = {Criswell, Alexander W. and Rieck, Steven and Mandic, Vuk},
  journal = {Phys. Rev. D},
  volume = {111},
  issue = {2},
  pages = {023025},
  numpages = {16},
  year = {2025},
  month = {Jan},
  publisher = {American Physical Society},
  doi = {10.1103/PhysRevD.111.023025},
  url = {https://link.aps.org/doi/10.1103/PhysRevD.111.023025}
}

@ARTICLE{2021MNRAS.500.4958W,
       author = {{Wilhelm}, Maite J.~C. and {Korol}, Valeriya and {Rossi}, Elena M. and {D'Onghia}, Elena},
        title = "{The Milky Way's bar structural properties from gravitational waves}",
      journal = {mnras},
         year = 2021,
        month = jan,
       volume = {500},
       number = {4},
        pages = {4958-4971},
          doi = {10.1093/mnras/staa3457},
archivePrefix = {arXiv},
       eprint = {2003.11074},
 primaryClass = {astro-ph.GA},
       adsurl = {https://ui.adsabs.harvard.edu/abs/2021MNRAS.500.4958W}
}

@ARTICLE{2019MNRAS.483.5518K,
       author = {{Korol}, Valeriya and {Rossi}, Elena M. and {Barausse}, Enrico},
        title = "{A multimessenger study of the Milky Way's stellar disc and bulge with LISA, Gaia, and LSST}",
      journal = {mnras},
         year = 2019,
        month = mar,
       volume = {483},
       number = {4},
        pages = {5518-5533},
          doi = {10.1093/mnras/sty3440},
archivePrefix = {arXiv},
       eprint = {1806.03306},
 primaryClass = {astro-ph.GA},
       adsurl = {https://ui.adsabs.harvard.edu/abs/2019MNRAS.483.5518K}
}

@software{katie_breivik_2019_3482915,
  author       = {Katie Breivik and
                  Scott Coughlin and
                  Michael Zevin and
                  Carl Rodriguez and
                  Jeff Andrews and
                  Chase Kimball and
                  mcdigman and
                  1nhtran},
  title        = {COSMIC-PopSynth/COSMIC: COSMIC now integrated with
                   Globular Cluster Simulation Software
                   ClusterMonteCarlo (CMC)
                  },
  month        = oct,
  year         = 2019,
  publisher    = {Zenodo},
  version      = {v3.1.0},
  doi          = {10.5281/zenodo.3482915},
  url          = {https://doi.org/10.5281/zenodo.3482915},
}

@misc{colpi_lisa_2024,
    title = {{LISA} {Definition} {Study} {Report}},
    url = {http://arxiv.org/abs/2402.07571},
    doi = {10.48550/arXiv.2402.07571},
    urldate = {2024-11-18},
    publisher = {arXiv},
    author = {Colpi, Monica and Danzmann, Karsten and Hewitson, Martin and Holley-Bockelmann, Kelly and Jetzer, Philippe and Nelemans, Gijs and Petiteau, Antoine and Shoemaker, David and Sopuerta, Carlos and Stebbins, Robin and Tanvir, Nial and Ward, Henry and Weber, William Joseph and Thorpe, Ira and Daurskikh, Anna and Deep, Atul and Núñez, Ignacio Fernández and Marirrodriga, César García and Gehler, Martin and Halain, Jean-Philippe and Jennrich, Oliver and Lammers, Uwe and Larrañaga, Jonan and Lieser, Maike and Lützgendorf, Nora and Martens, Waldemar and Mondin, Linda and Niño, Ana Piris and Amaro-Seoane, Pau and Sedda, Manuel Arca and Auclair, Pierre and Babak, Stanislav and Baghi, Quentin and Baibhav, Vishal and Baker, Tessa and Bayle, Jean-Baptiste and Berry, Christopher and Berti, Emanuele and Boileau, Guillaume and Bonetti, Matteo and Brito, Richard and Buscicchio, Riccardo and Calcagni, Gianluca and Capelo, Pedro R. and Caprini, Chiara and Caputo, Andrea and Castelli, Eleonora and Chen, Hsin-Yu and Chen, Xian and Chua, Alvin and Davies, Gareth and Derdzinski, Andrea and Domcke, Valerie Fiona and Doneva, Daniela and Dvorkin, Irna and Ezquiaga, Jose María and Gair, Jonathan and Haiman, Zoltan and Harry, Ian and Hartwig, Olaf and Hees, Aurelien and Heffernan, Anna and Husa, Sascha and Izquierdo, David and Karnesis, Nikolaos and Klein, Antoine and Korol, Valeriya and Korsakova, Natalia and Kupfer, Thomas and Laghi, Danny and Lamberts, Astrid and Larson, Shane and Jeune, Maude Le and Lewicki, Marek and Littenberg, Tyson and Madge, Eric and Mangiagli, Alberto and Marsat, Sylvain and Vilchez, Ivan Martin and Maselli, Andrea and Mathews, Josh and Meent, Maarten van de and Muratore, Martina and Nardini, Germano and Pani, Paolo and Peloso, Marco and Pieroni, Mauro and Pound, Adam and Quelquejay-Leclere, Hippolyte and Ricciardone, Angelo and Rossi, Elena Maria and Sartirana, Andrea and Savalle, Etienne and Sberna, Laura and Sesana, Alberto and Shoemaker, Deirdre and Slutsky, Jacob and Sotiriou, Thomas and Speri, Lorenzo and Staab, Martin and Steer, Danièle and Tamanini, Nicola and Tasinato, Gianmassimo and Torrado, Jesus and Torres-Orjuela, Alejandro and Toubiana, Alexandre and Vallisneri, Michele and Vecchio, Alberto and Volonteri, Marta and Yagi, Kent and Zwick, Lorenz},
    month = feb,
    year = {2024},
    note = {arXiv:2402.07571},
}

@article{astropy:2013,
        Adsurl = {https://adsabs.harvard.edu/abs/2013A%26A...558A..33A},
        Archiveprefix = {arXiv},
        Author = {{Astropy Collaboration} and {Robitaille}, T.~P. and {Tollerud}, E.~J. and {Greenfield}, P. and {Droettboom}, M. and {Bray}, E. and {Aldcroft}, T. and {Davis}, M. and {Ginsburg}, A. and {Price-Whelan}, A.~M. and {Kerzendorf}, W.~E. and {Conley}, A. and {Crighton}, N. and {Barbary}, K. and {Muna}, D. and {Ferguson}, H. and {Grollier}, F. and {Parikh}, M.~M. and {Nair}, P.~H. and {Unther}, H.~M. and {Deil}, C. and {Woillez}, J. and {Conseil}, S. and {Kramer}, R. and {Turner}, J.~E.~H. and {Singer}, L. and {Fox}, R. and {Weaver}, B.~A. and {Zabalza}, V. and {Edwards}, Z.~I. and {Azalee Bostroem}, K. and {Burke}, D.~J. and {Casey}, A.~R. and {Crawford}, S.~M. and {Dencheva}, N. and {Ely}, J. and {Jenness}, T. and {Labrie}, K. and {Lim}, P.~L. and {Pierfederici}, F. and {Pontzen}, A. and {Ptak}, A. and {Refsdal}, B. and {Servillat}, M. and {Streicher}, O.},
        Doi = {10.1051/0004-6361/201322068},
        Eid = {A33},
        Eprint = {1307.6212},
        Journal = {aap},
        Month = oct,
        Pages = {A33},
        Primaryclass = {astro-ph.IM},
        Title = {{Astropy: A community Python package for astronomy}},
        Volume = 558,
        Year = 2013}

@ARTICLE{astropy:2018,
               author = {{Astropy Collaboration} and {Price-Whelan}, A.~M. and
                 {Sip{\H{o}}cz}, B.~M. and {G{\"u}nther}, H.~M. and {Lim}, P.~L. and
                 {Crawford}, S.~M. and {Conseil}, S. and {Shupe}, D.~L. and
                 {Craig}, M.~W. and {Dencheva}, N. and {Ginsburg}, A. and {Vand
                erPlas}, J.~T. and {Bradley}, L.~D. and {P{\'e}rez-Su{\'a}rez}, D. and
                 {de Val-Borro}, M. and {Aldcroft}, T.~L. and {Cruz}, K.~L. and
                 {Robitaille}, T.~P. and {Tollerud}, E.~J. and {Ardelean}, C. and
                 {Babej}, T. and {Bach}, Y.~P. and {Bachetti}, M. and {Bakanov}, A.~V. and
                 {Bamford}, S.~P. and {Barentsen}, G. and {Barmby}, P. and
                 {Baumbach}, A. and {Berry}, K.~L. and {Biscani}, F. and {Boquien}, M. and
                 {Bostroem}, K.~A. and {Bouma}, L.~G. and {Brammer}, G.~B. and
                 {Bray}, E.~M. and {Breytenbach}, H. and {Buddelmeijer}, H. and
                 {Burke}, D.~J. and {Calderone}, G. and {Cano Rodr{\'\i}guez}, J.~L. and
                 {Cara}, M. and {Cardoso}, J.~V.~M. and {Cheedella}, S. and {Copin}, Y. and
                 {Corrales}, L. and {Crichton}, D. and {D'Avella}, D. and {Deil}, C. and
                 {Depagne}, {\'E}. and {Dietrich}, J.~P. and {Donath}, A. and
                 {Droettboom}, M. and {Earl}, N. and {Erben}, T. and {Fabbro}, S. and
                 {Ferreira}, L.~A. and {Finethy}, T. and {Fox}, R.~T. and
                 {Garrison}, L.~H. and {Gibbons}, S.~L.~J. and {Goldstein}, D.~A. and
                 {Gommers}, R. and {Greco}, J.~P. and {Greenfield}, P. and
                 {Groener}, A.~M. and {Grollier}, F. and {Hagen}, A. and {Hirst}, P. and
                 {Homeier}, D. and {Horton}, A.~J. and {Hosseinzadeh}, G. and {Hu}, L. and
                 {Hunkeler}, J.~S. and {Ivezi{\'c}}, {\v{Z}}. and {Jain}, A. and
                 {Jenness}, T. and {Kanarek}, G. and {Kendrew}, S. and {Kern}, N.~S. and
                 {Kerzendorf}, W.~E. and {Khvalko}, A. and {King}, J. and {Kirkby}, D. and
                 {Kulkarni}, A.~M. and {Kumar}, A. and {Lee}, A. and {Lenz}, D. and
                 {Littlefair}, S.~P. and {Ma}, Z. and {Macleod}, D.~M. and
                 {Mastropietro}, M. and {McCully}, C. and {Montagnac}, S. and
                 {Morris}, B.~M. and {Mueller}, M. and {Mumford}, S.~J. and {Muna}, D. and
                 {Murphy}, N.~A. and {Nelson}, S. and {Nguyen}, G.~H. and
                 {Ninan}, J.~P. and {N{\"o}the}, M. and {Ogaz}, S. and {Oh}, S. and
                 {Parejko}, J.~K. and {Parley}, N. and {Pascual}, S. and {Patil}, R. and
                 {Patil}, A.~A. and {Plunkett}, A.~L. and {Prochaska}, J.~X. and
                 {Rastogi}, T. and {Reddy Janga}, V. and {Sabater}, J. and
                 {Sakurikar}, P. and {Seifert}, M. and {Sherbert}, L.~E. and
                 {Sherwood-Taylor}, H. and {Shih}, A.~Y. and {Sick}, J. and
                 {Silbiger}, M.~T. and {Singanamalla}, S. and {Singer}, L.~P. and
                 {Sladen}, P.~H. and {Sooley}, K.~A. and {Sornarajah}, S. and
                 {Streicher}, O. and {Teuben}, P. and {Thomas}, S.~W. and
                 {Tremblay}, G.~R. and {Turner}, J.~E.~H. and {Terr{\'o}n}, V. and
                 {van Kerkwijk}, M.~H. and {de la Vega}, A. and {Watkins}, L.~L. and
                 {Weaver}, B.~A. and {Whitmore}, J.~B. and {Woillez}, J. and
                 {Zabalza}, V. and {Astropy Contributors}},
                 title = "{The Astropy Project: Building an Open-science Project and Status of the v2.0 Core Package}",
                 journal = {aj},
                 year = 2018,
                month = sep,
               volume = {156},
               number = {3},
                  eid = {123},
                pages = {123},
                  doi = {10.3847/1538-3881/aabc4f},
        archivePrefix = {arXiv},
               eprint = {1801.02634},
         primaryClass = {astro-ph.IM},
               adsurl = {https://ui.adsabs.harvard.edu/abs/2018AJ....156..123A}
        }

@ARTICLE{astropy:2022,
               author = {{Astropy Collaboration} and {Price-Whelan}, Adrian M. and {Lim}, Pey Lian and {Earl}, Nicholas and {Starkman}, Nathaniel and {Bradley}, Larry and {Shupe}, David L. and {Patil}, Aarya A. and {Corrales}, Lia and {Brasseur}, C.~E. and {N{\"o}the}, Maximilian and {Donath}, Axel and {Tollerud}, Erik and {Morris}, Brett M. and {Ginsburg}, Adam and {Vaher}, Eero and {Weaver}, Benjamin A. and {Tocknell}, James and {Jamieson}, William and {van Kerkwijk}, Marten H. and {Robitaille}, Thomas P. and {Merry}, Bruce and {Bachetti}, Matteo and {G{\"u}nther}, H. Moritz and {Aldcroft}, Thomas L. and {Alvarado-Montes}, Jaime A. and {Archibald}, Anne M. and {B{\'o}di}, Attila and {Bapat}, Shreyas and {Barentsen}, Geert and {Baz{\'a}n}, Juanjo and {Biswas}, Manish and {Boquien}, M{\'e}d{\'e}ric and {Burke}, D.~J. and {Cara}, Daria and {Cara}, Mihai and {Conroy}, Kyle E. and {Conseil}, Simon and {Craig}, Matthew W. and {Cross}, Robert M. and {Cruz}, Kelle L. and {D'Eugenio}, Francesco and {Dencheva}, Nadia and {Devillepoix}, Hadrien A.~R. and {Dietrich}, J{\"o}rg P. and {Eigenbrot}, Arthur Davis and {Erben}, Thomas and {Ferreira}, Leonardo and {Foreman-Mackey}, Daniel and {Fox}, Ryan and {Freij}, Nabil and {Garg}, Suyog and {Geda}, Robel and {Glattly}, Lauren and {Gondhalekar}, Yash and {Gordon}, Karl D. and {Grant}, David and {Greenfield}, Perry and {Groener}, Austen M. and {Guest}, Steve and {Gurovich}, Sebastian and {Handberg}, Rasmus and {Hart}, Akeem and {Hatfield-Dodds}, Zac and {Homeier}, Derek and {Hosseinzadeh}, Griffin and {Jenness}, Tim and {Jones}, Craig K. and {Joseph}, Prajwel and {Kalmbach}, J. Bryce and {Karamehmetoglu}, Emir and {Ka{\l}uszy{\'n}ski}, Miko{\l}aj and {Kelley}, Michael S.~P. and {Kern}, Nicholas and {Kerzendorf}, Wolfgang E. and {Koch}, Eric W. and {Kulumani}, Shankar and {Lee}, Antony and {Ly}, Chun and {Ma}, Zhiyuan and {MacBride}, Conor and {Maljaars}, Jakob M. and {Muna}, Demitri and {Murphy}, N.~A. and {Norman}, Henrik and {O'Steen}, Richard and {Oman}, Kyle A. and {Pacifici}, Camilla and {Pascual}, Sergio and {Pascual-Granado}, J. and {Patil}, Rohit R. and {Perren}, Gabriel I. and {Pickering}, Timothy E. and {Rastogi}, Tanuj and {Roulston}, Benjamin R. and {Ryan}, Daniel F. and {Rykoff}, Eli S. and {Sabater}, Jose and {Sakurikar}, Parikshit and {Salgado}, Jes{\'u}s and {Sanghi}, Aniket and {Saunders}, Nicholas and {Savchenko}, Volodymyr and {Schwardt}, Ludwig and {Seifert-Eckert}, Michael and {Shih}, Albert Y. and {Jain}, Anany Shrey and {Shukla}, Gyanendra and {Sick}, Jonathan and {Simpson}, Chris and {Singanamalla}, Sudheesh and {Singer}, Leo P. and {Singhal}, Jaladh and {Sinha}, Manodeep and {Sip{\H{o}}cz}, Brigitta M. and {Spitler}, Lee R. and {Stansby}, David and {Streicher}, Ole and {{\v{S}}umak}, Jani and {Swinbank}, John D. and {Taranu}, Dan S. and {Tewary}, Nikita and {Tremblay}, Grant R. and {Val-Borro}, Miguel de and {Van Kooten}, Samuel J. and {Vasovi{\'c}}, Zlatan and {Verma}, Shresth and {de Miranda Cardoso}, Jos{\'e} Vin{\'i}cius and {Williams}, Peter K.~G. and {Wilson}, Tom J. and {Winkel}, Benjamin and {Wood-Vasey}, W.~M. and {Xue}, Rui and {Yoachim}, Peter and {Zhang}, Chen and {Zonca}, Andrea and {Astropy Project Contributors}},
                title = "{The Astropy Project: Sustaining and Growing a Community-oriented Open-source Project and the Latest Major Release (v5.0) of the Core Package}",
              journal = {\apj},
                 year = 2022,
                month = aug,
               volume = {935},
               number = {2},
                  eid = {167},
                pages = {167},
                  doi = {10.3847/1538-4357/ac7c74},
        archivePrefix = {arXiv},
               eprint = {2206.14220},
         primaryClass = {astro-ph.IM},
               adsurl = {https://ui.adsabs.harvard.edu/abs/2022ApJ...935..167A}
        }

@article{robson_construction_2019,
    title = {The construction and use of {LISA} sensitivity curves},
    volume = {36},
    issn = {0264-9381, 1361-6382},
    url = {http://arxiv.org/abs/1803.01944},
    doi = {10.1088/1361-6382/ab1101},
    number = {10},
    urldate = {2026-03-02},
    journal = {Classical and Quantum Gravity},
    author = {Robson, Travis and Cornish, Neil and Liu, Chang},
    month = may,
    year = {2019},
    note = {arXiv:1803.01944 [astro-ph]},
    pages = {105011},
}

@article{Zonca2019,
  doi = {10.21105/joss.01298},
  url = {https://doi.org/10.21105/joss.01298},
  year = {2019},
  month = mar,
  publisher = {The Open Journal},
  volume = {4},
  number = {35},
  pages = {1298},
  author = {Andrea Zonca and Leo Singer and Daniel Lenz and Martin Reinecke and Cyrille Rosset and Eric Hivon and Krzysztof Gorski},
  title = {healpy: equal area pixelization and spherical harmonics transforms for data on the sphere in Python},
  journal = {Journal of Open Source Software}
}

@ARTICLE{2005ApJ...622..759G,
   author = {{G{\'o}rski}, K.~M. and {Hivon}, E. and {Banday}, A.~J. and 
	{Wandelt}, B.~D. and {Hansen}, F.~K. and {Reinecke}, M. and 
	{Bartelmann}, M.},
    title = "{HEALPix: A Framework for High-Resolution Discretization and Fast Analysis of Data Distributed on the Sphere}",
  journal = {\apj},
   eprint = {arXiv:astro-ph/0409513},
     year = 2005,
    month = apr,
   volume = 622,
    pages = {759-771},
      doi = {10.1086/427976},
   adsurl = {http://adsabs.harvard.edu/abs/2005ApJ...622..759G}
}

@article{littenberg_prototype_2023,
    title = {Prototype global analysis of {LISA} data with multiple source types},
    volume = {107},
    issn = {1550-79980556-2821},
    url = {https://ui.adsabs.harvard.edu/abs/2023PhRvD.107f3004L},
    doi = {10.1103/PhysRevD.107.063004},
    urldate = {2024-11-08},
    journal = {Physical Review D},
    publisher = {APS},
    author = {Littenberg, Tyson B. and Cornish, Neil J.},
    month = mar,
    year = {2023},
    note = {ADS Bibcode: 2023PhRvD.107f3004L},
    pages = {063004},
}

@ARTICLE{2025PhRvD.111b4060K,
       author = {{Katz}, Michael L. and {Karnesis}, Nikolaos and {Korsakova}, Natalia and {Gair}, Jonathan R. and {Stergioulas}, Nikolaos},
        title = "{Efficient GPU-accelerated multisource global fit pipeline for LISA data analysis}",
      journal = {\prd},
         year = 2025,
        month = jan,
       volume = {111},
       number = {2},
          eid = {024060},
        pages = {024060},
          doi = {10.1103/PhysRevD.111.024060},
archivePrefix = {arXiv},
       eprint = {2405.04690},
 primaryClass = {gr-qc},
       adsurl = {https://ui.adsabs.harvard.edu/abs/2025PhRvD.111b4060K}
}

@article{romano2017detection,
  title={Detection methods for stochastic gravitational-wave backgrounds: a unified treatment},
  author={Romano, Joseph D and Cornish, Neil J},
  journal={Living reviews in relativity},
  volume={20},
  number={1},
  pages={2},
  year={2017},
  publisher={Springer}
}

@article{cornish_space_2001,
    title = {Space missions to detect the cosmic gravitational-wave background},
    volume = {18},
    issn = {0264-9381},
    url = {https://ui.adsabs.harvard.edu/abs/2001CQGra..18.3473C},
    doi = {10.1088/0264-9381/18/17/308},
    urldate = {2026-05-18},
    journal = {Classical and Quantum Gravity},
    publisher = {IOP},
    author = {Cornish, Neil J. and Larson, Shane L.},
    month = sep,
    year = {2001},
    note = {ADS Bibcode: 2001CQGra..18.3473C},
    pages = {3473--3495},
}

@article{PhysRevD.65.022004,
  title = {Detecting a stochastic gravitational wave background with the Laser Interferometer Space Antenna},
  author = {Cornish, Neil J.},
  journal = {Phys. Rev. D},
  volume = {65},
  issue = {2},
  pages = {022004},
  numpages = {11},
  year = {2001},
  month = {Dec},
  publisher = {American Physical Society},
  doi = {10.1103/PhysRevD.65.022004},
  url = {https://link.aps.org/doi/10.1103/PhysRevD.65.022004}
}

@article{banagiri_mapping_2021,
    title = {Mapping the gravitational-wave sky with {LISA}: a {Bayesian} spherical harmonic approach},
    volume = {507},
    issn = {0035-8711},
    shorttitle = {Mapping the gravitational-wave sky with {LISA}},
    url = {https://doi.org/10.1093/mnras/stab2479},
    doi = {10.1093/mnras/stab2479},
    number = {4},
    urldate = {2025-04-28},
    journal = {Monthly Notices of the Royal Astronomical Society},
    author = {Banagiri, Sharan and Criswell, Alexander and Kuan, Tommy and Mandic, Vuk and Romano, Joseph D and Taylor, Stephen R},
    month = nov,
    year = {2021},
    pages = {5451--5462},
}

@article{collaboration2019gaia,
  title={Gaia Data Release 2},
  author={Collaboration, Gaia and Eyer, L and Rimoldini, L and Audard, M and Anderson, RI and Nienartowicz, K and Glass, F and Marchal, O and Grenon, M and Mowlavi, N and others},
  journal={A\&A},
  volume={623},
  pages={A110},
  year={2019}
}

@article{collaboration2018gaia,
  title={Gaia Data Release 2},
  author={Collaboration, Gaia and Mignard, F and Klioner, SA and Lindegren, L and Hern{\'a}ndez, J and Bastian, U and Bombrun, A and Hobbs, D and Lammers, U and Michalik, D and others},
  journal={A\&A},
  volume={616},
  pages={A14},
  year={2018}
}

@article{amaro-seoane_astrophysics_2023,
    title = {Astrophysics with the {Laser} {Interferometer} {Space} {Antenna}},
    volume = {26},
    issn = {1433-8351},
    url = {https://doi.org/10.1007/s41114-022-00041-y},
    doi = {10.1007/s41114-022-00041-y},
    number = {1},
    urldate = {2026-07-01},
    journal = {Living Reviews in Relativity},
    author = {Amaro-Seoane, Pau and Andrews, Jeff and Arca Sedda, Manuel and Askar, Abbas and Baghi, Quentin and Balasov, Razvan and Bartos, Imre and Bavera, Simone S. and Bellovary, Jillian and Berry, Christopher P. L. and Berti, Emanuele and Bianchi, Stefano and Blecha, Laura and Blondin, {Stéphane} and {Bogdanović}, Tamara and Boissier, Samuel and Bonetti, Matteo and Bonoli, Silvia and Bortolas, Elisa and Breivik, Katelyn and Capelo, Pedro R. and Caramete, Laurentiu and Cattorini, Federico and Charisi, Maria and Chaty, Sylvain and Chen, Xian and {Chruślińska}, Martyna and Chua, Alvin J. K. and Church, Ross and Colpi, Monica and D’Orazio, Daniel and Danielski, Camilla and Davies, Melvyn B. and Dayal, Pratika and De Rosa, Alessandra and Derdzinski, Andrea and Destounis, Kyriakos and Dotti, Massimo and Duţan, Ioana and Dvorkin, Irina and Fabj, Gaia and Foglizzo, Thierry and Ford, Saavik and Fouvry, Jean-Baptiste and Franchini, Alessia and Fragos, Tassos and Fryer, Chris and Gaspari, Massimo and Gerosa, Davide and Graziani, Luca and Groot, Paul and Habouzit, Melanie and Haggard, Daryl and Haiman, Zoltan and Han, Wen-Biao and Istrate, Alina and Johansson, Peter H. and Khan, Fazeel Mahmood and Kimpson, Tomas and Kokkotas, Kostas and Kong, Albert and Korol, Valeriya and Kremer, Kyle and Kupfer, Thomas and Lamberts, Astrid and Larson, Shane and Lau, Mike and Liu, Dongliang and Lloyd-Ronning, Nicole and Lodato, Giuseppe and Lupi, Alessandro and Ma, Chung-Pei and Maccarone, Tomas and Mandel, Ilya and Mangiagli, Alberto and Mapelli, Michela and Mathis, Stéphane and Mayer, Lucio and McGee, Sean and McKernan, Berry and Miller, M. Coleman and Mota, David F. and Mumpower, Matthew and Nasim, Syeda S. and Nelemans, Gijs and Noble, Scott and Pacucci, Fabio and Panessa, Francesca and Paschalidis, Vasileios and Pfister, Hugo and Porquet, Delphine and Quenby, John and Ricarte, Angelo and Röpke, Friedrich K. and Regan, John and Rosswog, Stephan and Ruiter, Ashley and Ruiz, Milton and Runnoe, Jessie and Schneider, Raffaella and Schnittman, Jeremy and Secunda, Amy and Sesana, Alberto and Seto, Naoki and Shao, Lijing and Shapiro, Stuart and Sopuerta, Carlos and Stone, Nicholas C. and Suvorov, Arthur and Tamanini, Nicola and Tamfal, Tomas and Tauris, Thomas and Temmink, Karel and Tomsick, John and Toonen, Silvia and Torres-Orjuela, Alejandro and Toscani, Martina and Tsokaros, Antonios and Unal, Caner and {Vázquez-Aceves}, {Verónica} and Valiante, Rosa and van Putten, Maurice and van Roestel, Jan and Vignali, Christian and Volonteri, Marta and Wu, Kinwah and Younsi, Ziri and Yu, Shenghua and Zane, Silvia and Zwick, Lorenz and Antonini, Fabio and Baibhav, Vishal and Barausse, Enrico and Bonilla Rivera, Alexander and Branchesi, Marica and Branduardi-Raymont, Graziella and Burdge, Kevin and Chakraborty, Srija and Cuadra, Jorge and Dage, Kristen and Davis, Benjamin and de Mink, Selma E. and Decarli, Roberto and Doneva, Daniela and Escoffier, Stephanie and Gandhi, Poshak and Haardt, Francesco and Lousto, Carlos O. and Nissanke, Samaya and Nordhaus, Jason and O’Shaughnessy, Richard and Portegies Zwart, Simon and Pound, Adam and Schussler, Fabian and Sergijenko, Olga and Spallicci, Alessandro and Vernieri, Daniele and {Vigna-Gómez}, Alejandro},
    month = mar,
    year = {2023},
    pages = {2},
}

@article{korol_observationally_2022,
    title = {Observationally driven {Galactic} double white dwarf population for {LISA}},
    volume = {511},
    issn = {0035-8711},
    url = {https://ui.adsabs.harvard.edu/abs/2022MNRAS.511.5936K},
    doi = {10.1093/mnras/stac415},
    urldate = {2026-07-07},
    journal = {Monthly Notices of the Royal Astronomical Society},
    publisher = {OUP},
    author = {Korol, Valeriya and Hallakoun, Na'ama and Toonen, Silvia and Karnesis, Nikolaos},
    month = apr,
    year = {2022},
    note = {ADS Bibcode: 2022MNRAS.511.5936K},
    pages = {5936--5947},
}

@book{tauris_physics_2022,
    address = {Princeton},
    edition = {1st},
    series = {Princeton series in astrophysics},
    title = {Physics of binary star evolution: from stars to x-ray binaries and gravitational wave sources},
    isbn = {978-0-691-17907-0 978-0-691-17908-7},
    shorttitle = {Physics of binary star evolution},
    publisher = {Princeton University Press},
    author = {Tauris, Thomas M. and van den Heuvel, Edward P. J.},
    year = {2022},
}

@book{warner2003cataclysmic,
  title={Cataclysmic variable stars},
  author={Warner, Brian},
  volume={28},
  year={2003},
  publisher={Cambridge University Press}
}

@article{rodriguez_cataclysmic_2025,
    title = {Cataclysmic {Variables} and {AM} {CVn} {Binaries} in {SRG}/{eROSITA} + {Gaia}: {Volume} {Limited} {Samples}, {X}-{Ray} {Luminosity} {Functions}, and {Space} {Densities}},
    volume = {137},
    issn = {0004-6280},
    shorttitle = {Cataclysmic {Variables} and {AM} {CVn} {Binaries} in {SRG}/{eROSITA} + {Gaia}},
    url = {https://ui.adsabs.harvard.edu/abs/2025PASP..137a4201R},
    doi = {10.1088/1538-3873/ada185},
    urldate = {2026-06-08},
    journal = {Publications of the Astronomical Society of the Pacific},
    publisher = {IOP},
    author = {Rodriguez, Antonio C. and El-Badry, Kareem and Suleimanov, Valery and Pala, Anna F. and Kulkarni, Shrinivas R. and Gaensicke, Boris and Mori, Kaya and Rich, R. Michael and Sarkar, Arnab and Bao, Tong and Lopes de Oliveira, Raimundo and Ramsay, Gavin and Szkody, Paula and Graham, Matthew and Prince, Thomas A. and Caiazzo, Ilaria and Vanderbosch, Zachary P. and van Roestel, Jan and Das, Kaustav K. and Qin, Yu-Jing and Kasliwal, Mansi M. and Wold, Avery and Groom, Steven L. and Reiley, Daniel and Riddle, Reed},
    month = jan,
    year = {2025},
    note = {ADS Bibcode: 2025PASP..137a4201R},
    pages = {014201},
}

@article{belloni_reversing_2023,
    title = {Reversing the verdict: {Cataclysmic} variables could be the dominant progenitors of {AM} {CVn} binaries after all},
    volume = {678},
    issn = {0004-6361},
    shorttitle = {Reversing the verdict},
    url = {https://ui.adsabs.harvard.edu/abs/2023A&A...678A..34B},
    doi = {10.1051/0004-6361/202347047},
    urldate = {2026-06-18},
    journal = {Astronomy and Astrophysics},
    publisher = {EDP},
    author = {Belloni, Diogo and Schreiber, Matthias R.},
    month = oct,
    year = {2023},
    note = {ADS Bibcode: 2023A\&A...678A..34B},
    pages = {A34},
}

@article{schreiber_three_2016,
    title = {Three in one go: consequential angular momentum loss can solve major problems of {CV} evolution},
    volume = {455},
    issn = {0035-8711},
    shorttitle = {Three in one go},
    url = {https://ui.adsabs.harvard.edu/abs/2016MNRAS.455L..16S},
    doi = {10.1093/mnrasl/slv144},
    urldate = {2026-08-18},
    journal = {Monthly Notices of the Royal Astronomical Society},
    publisher = {OUP},
    author = {Schreiber, M. R. and Zorotovic, M. and Wijnen, T. P. G.},
    month = jan,
    year = {2016},
    note = {ADS Bibcode: 2016MNRAS.455L..16S},
    pages = {L16--L20},
}

@article{zorotovic_cataclysmic_2020,
    title = {Cataclysmic variable evolution and the white dwarf mass problem: {A} {Review}},
    volume = {66},
    issn = {0273-1177},
    shorttitle = {Cataclysmic variable evolution and the white dwarf mass problem},
    url = {https://ui.adsabs.harvard.edu/abs/2020AdSpR..66.1080Z},
    doi = {10.1016/j.asr.2019.08.044},
    urldate = {2026-07-01},
    journal = {Advances in Space Research},
    publisher = {Elsevier},
    author = {Zorotovic, Monica and Schreiber, Matthias R.},
    month = sep,
    year = {2020},
    note = {ADS Bibcode: 2020AdSpR..66.1080Z},
    pages = {1080--1089},
}

@article{metzger_transients_2021,
    title = {Transients from the {Cataclysmic} {Deaths} of {Cataclysmic} {Variables}},
    volume = {923},
    issn = {0004-637X},
    url = {https://ui.adsabs.harvard.edu/abs/2021ApJ...923..100M},
    doi = {10.3847/1538-4357/ac2a39},
    urldate = {2026-08-18},
    journal = {The Astrophysical Journal},
    publisher = {IOP},
    author = {Metzger, Brian D. and Zenati, Yossef and Chomiuk, Laura and Shen, Ken J. and Strader, Jay},
    month = dec,
    year = {2021},
    note = {ADS Bibcode: 2021ApJ...923..100M},
    pages = {100},
}

@misc{knigge_evolution_2011,
    title = {The {Evolution} of {Cataclysmic} {Variables}},
    url = {http://arxiv.org/abs/1108.4716},
    doi = {10.48550/arXiv.1108.4716},
    urldate = {2025-10-10},
    publisher = {arXiv},
    author = {Knigge, Christian},
    month = sep,
    year = {2011},
    note = {arXiv:1108.4716 [astro-ph]},
}

@article{barraza-jorquera_modest_2026,
    title = {A modest change in magnetic braking at the fully convective boundary explains the evolution of cataclysmic variables},
    volume = {708},
    issn = {0004-6361},
    url = {https://ui.adsabs.harvard.edu/abs/2026A&A...708L..11B},
    doi = {10.1051/0004-6361/202659326},
    urldate = {2026-08-25},
    journal = {Astronomy and Astrophysics},
    publisher = {EDP},
    author = {Barraza-Jorquera, Joaquín A. and Schreiber, Matthias R. and Littlefair, Stuart and Belloni, Diogo and Schwope, Axel D.},
    month = apr,
    year = {2026},
    note = {ADS Bibcode: 2026A\&A...708L..11B},
    pages = {L11},
}

@ARTICLE{1983ApJ...275..713R,
       author = {{Rappaport}, S. and {Verbunt}, F. and {Joss}, P.~C.},
        title = "{A new technique for calculations of binary stellar evolution application to magnetic braking.}",
      journal = {\apj},
         year = 1983,
        month = dec,
       volume = {275},
        pages = {713-731},
          doi = {10.1086/161569},
       adsurl = {https://ui.adsabs.harvard.edu/abs/1983ApJ...275..713R}
}

@ARTICLE{1983A&A...124..267S,
       author = {{Spruit}, H.~C. and {Ritter}, H.},
        title = "{Stellar activity and the period gap in cataclysmic variables}",
      journal = {aap},
         year = 1983,
        month = aug,
       volume = {124},
       number = {2},
        pages = {267-272},
       adsurl = {https://ui.adsabs.harvard.edu/abs/1983A&A...124..267S}
}

@article{schreiber_cataclysmic_2024,
    title = {The cataclysmic variable orbital period gap: {More} evident than ever},
    volume = {682},
    issn = {0004-6361},
    shorttitle = {The cataclysmic variable orbital period gap},
    url = {https://ui.adsabs.harvard.edu/abs/2024A&A...682L...7S},
    doi = {10.1051/0004-6361/202348807},
    urldate = {2026-09-02},
    journal = {Astronomy and Astrophysics},
    publisher = {EDP},
    author = {Schreiber, Matthias R. and Belloni, Diogo and Schwope, Axel D.},
    month = feb,
    year = {2024},
    note = {ADS Bibcode: 2024A\&A...682L...7S},
    pages = {L7},
}

@article{zwart1996population,
  title={Population synthesis of high-mass binaries},
  author={Zwart, SF Portegies and Verbunt, Frank},
  journal={A\&a},
  volume={309},
  pages={79--196},
  year={1996}
}

@article{toonen2012supernova,
  title={Supernova Type Ia progenitors from merging double white dwarfs: Using a new population synthesis model},
  author={Toonen, Silvia and Nelemans, Gijs and Portegies Zwart, S},
  journal={Astronomy \& Astrophysics},
  volume={546},
  pages={A70},
  year={2012},
  publisher={EDP Sciences}
}

@ARTICLE{2001A&A...365..491N,
       author = {{Nelemans}, G. and {Yungelson}, L.~R. and {Portegies Zwart}, S.~F. and {Verbunt}, F.},
        title = "{Population synthesis for double white dwarfs . I. Close detached systems}",
      journal = {aap},
         year = 2001,
        month = jan,
       volume = {365},
        pages = {491-507},
          doi = {10.1051/0004-6361:20000147},
archivePrefix = {arXiv},
       eprint = {astro-ph/0010457},
 primaryClass = {astro-ph},
       adsurl = {https://ui.adsabs.harvard.edu/abs/2001A&A...365..491N}
}

@ARTICLE{2001A&A...368..939N,
       author = {{Nelemans}, G. and {Portegies Zwart}, S.~F. and {Verbunt}, F. and {Yungelson}, L.~R.},
        title = "{Population synthesis for double white dwarfs. II. Semi-detached systems: AM CVn stars}",
      journal = {aap},
         year = 2001,
        month = mar,
       volume = {368},
        pages = {939-949},
          doi = {10.1051/0004-6361:20010049},
archivePrefix = {arXiv},
       eprint = {astro-ph/0101123},
 primaryClass = {astro-ph},
       adsurl = {https://ui.adsabs.harvard.edu/abs/2001A&A...368..939N}
}

@ARTICLE{2001MNRAS.322..231K,
       author = {{Kroupa}, Pavel},
        title = "{On the variation of the initial mass function}",
      journal = {mnras},
         year = 2001,
        month = apr,
       volume = {322},
       number = {2},
        pages = {231-246},
          doi = {10.1046/j.1365-8711.2001.04022.x},
archivePrefix = {arXiv},
       eprint = {astro-ph/0009005},
 primaryClass = {astro-ph},
       adsurl = {https://ui.adsabs.harvard.edu/abs/2001MNRAS.322..231K}
}

@ARTICLE{2013ARA&A..51..269D,
       author = {{Duch{\^e}ne}, Gaspard and {Kraus}, Adam},
        title = "{Stellar Multiplicity}",
      journal = {araa},
         year = 2013,
        month = aug,
       volume = {51},
       number = {1},
        pages = {269-310},
          doi = {10.1146/annurev-astro-081710-102602},
archivePrefix = {arXiv},
       eprint = {1303.3028},
 primaryClass = {astro-ph.SR},
       adsurl = {https://ui.adsabs.harvard.edu/abs/2013ARA&A..51..269D}
}

@ARTICLE{1964PhRv..136.1224P,
       author = {{Peters}, P.~C.},
        title = "{Gravitational Radiation and the Motion of Two Point Masses}",
      journal = {Physical Review},
         year = 1964,
        month = nov,
       volume = {136},
       number = {4B},
        pages = {1224-1232},
          doi = {10.1103/PhysRev.136.B1224},
       adsurl = {https://ui.adsabs.harvard.edu/abs/1964PhRv..136.1224P}
}

@ARTICLE{1999MNRAS.307..857B,
       author = {{Boissier}, S. and {Prantzos}, N.},
        title = "{Chemo-spectrophotometric evolution of spiral galaxies - I. The model and the Milky Way}",
      journal = {mnras},
         year = 1999,
        month = aug,
       volume = {307},
       number = {4},
        pages = {857-876},
          doi = {10.1046/j.1365-8711.1999.02699.x},
archivePrefix = {arXiv},
       eprint = {astro-ph/9902148},
 primaryClass = {astro-ph},
       adsurl = {https://ui.adsabs.harvard.edu/abs/1999MNRAS.307..857B}
}

@ARTICLE{2020ApJ...890..117D,
       author = {{D'Onghia}, Elena and {L. Aguerri}, J. Alfonso},
        title = "{Trojans in the Solar Neighborhood}",
      journal = {\apj},
         year = 2020,
        month = feb,
       volume = {890},
       number = {2},
          eid = {117},
        pages = {117},
          doi = {10.3847/1538-4357/ab6bd6},
archivePrefix = {arXiv},
       eprint = {1907.08484},
 primaryClass = {astro-ph.GA},
       adsurl = {https://ui.adsabs.harvard.edu/abs/2020ApJ...890..117D}
}

@ARTICLE{2021MNRAS.507.3593M,
       author = {{Moe}, Maxwell and {Kratter}, Kaitlin M.},
        title = "{Impact of binary stars on planet statistics - I. Planet occurrence rates and trends with stellar mass}",
      journal = {mnras},
         year = 2021,
        month = nov,
       volume = {507},
       number = {3},
        pages = {3593-3611},
          doi = {10.1093/mnras/stab2328},
archivePrefix = {arXiv},
       eprint = {1912.01699},
 primaryClass = {astro-ph.EP},
       adsurl = {https://ui.adsabs.harvard.edu/abs/2021MNRAS.507.3593M}
}

@ARTICLE{1992ApJ...401..265M,
       author = {{Mazeh}, Tsevi and {Goldberg}, Dorit and {Duquennoy}, Antoine and {Mayor}, Michel},
        title = "{On the Mass-Ratio Distribution of Spectroscopic Binaries with Solar-Type Primaries}",
      journal = {\apj},
         year = 1992,
        month = dec,
       volume = {401},
        pages = {265},
          doi = {10.1086/172058},
       adsurl = {https://ui.adsabs.harvard.edu/abs/1992ApJ...401..265M}
}

@ARTICLE{1994A&A...282..801G,
       author = {{Goldberg}, D. and {Mazeh}, T.},
        title = "{The mass-ratio distribution of the spectroscopic binaries in the Pleiades.}",
      journal = {aap},
         year = 1994,
        month = feb,
       volume = {282},
        pages = {801-803},
       adsurl = {https://ui.adsabs.harvard.edu/abs/1994A&A...282..801G}
}

@ARTICLE{2019ApJ...872..165G,
       author = {{Geller}, Aaron M. and {Leigh}, Nathan W.~C. and {Giersz}, Mirek and {Kremer}, Kyle and {Rasio}, Frederic A.},
        title = "{In Search of the Thermal Eccentricity Distribution}",
      journal = {\apj},
         year = 2019,
        month = feb,
       volume = {872},
       number = {2},
          eid = {165},
        pages = {165},
          doi = {10.3847/1538-4357/ab0214},
archivePrefix = {arXiv},
       eprint = {1902.00019},
 primaryClass = {astro-ph.SR},
       adsurl = {https://ui.adsabs.harvard.edu/abs/2019ApJ...872..165G}
}

@book{maggiore_gravitational_2017,
    address = {Oxford},
    title = {Gravitational waves {Vol2}},
    isbn = {978-0-19-875528-9 978-0-19-857074-5 978-0-19-857089-9},
    publisher = {Oxford university press},
    author = {Maggiore, Michele},
    year = {2017},
}

@article{rieck_stochastic_2024,
    title = {A stochastic gravitational wave background in {LISA} from unresolved white dwarf binaries in the {Large} {Magellanic} {Cloud}},
    volume = {531},
    issn = {0035-8711},
    url = {https://ui.adsabs.harvard.edu/abs/2024MNRAS.531.2642R},
    doi = {10.1093/mnras/stae1283},
    urldate = {2026-05-18},
    journal = {Monthly Notices of the Royal Astronomical Society},
    publisher = {OUP},
    author = {Rieck, Steven and Criswell, Alexander W. and Korol, Valeriya and Keim, Michael A. and Bloom, Malachy and Mandic, Vuk},
    month = jun,
    year = {2024},
    note = {ADS Bibcode: 2024MNRAS.531.2642R},
    pages = {2642--2652},
}

@ARTICLE{1999PhRvD..59j2003T,
       author = {{Tinto}, Massimo and {Armstrong}, J.~W.},
        title = "{Cancellation of laser noise in an unequal-arm interferometer detector of gravitational radiation}",
      journal = {\prd},
         year = 1999,
        month = may,
       volume = {59},
       number = {10},
          eid = {102003},
        pages = {102003},
          doi = {10.1103/PhysRevD.59.102003},
       adsurl = {https://ui.adsabs.harvard.edu/abs/1999PhRvD..59j2003T}
}

@ARTICLE{2000PhRvD..63b1101T,
       author = {{Tinto}, Massimo and {Armstrong}, J.~W. and {Estabrook}, F.~B.},
        title = "{Discriminating a gravitational wave background from instrumental noise in the LISA detector}",
      journal = {\prd},
         year = 2000,
        month = dec,
       volume = {63},
       number = {2},
          eid = {021101},
        pages = {021101},
          doi = {10.1103/PhysRevD.63.021101},
       adsurl = {https://ui.adsabs.harvard.edu/abs/2000PhRvD..63b1101T}
}

@article{adams_discriminating_2010,
    title = {Discriminating between a stochastic gravitational wave background and instrument noise},
    volume = {82},
    issn = {1550-79980556-2821},
    url = {https://ui.adsabs.harvard.edu/abs/2010PhRvD..82b2002A},
    doi = {10.1103/PhysRevD.82.022002},
    urldate = {2026-05-18},
    journal = {Physical Review D},
    publisher = {APS},
    author = {Adams, Matthew R. and Cornish, Neil J.},
    month = jul,
    year = {2010},
    note = {ADS Bibcode: 2010PhRvD..82b2002A},
    pages = {022002},
}

@article{phinney_practical_2001,
    title = {A {Practical} {Theorem} on {Gravitational} {Wave} {Backgrounds}},
    url = {https://ui.adsabs.harvard.edu/abs/2001astro.ph..8028P},
    doi = {10.48550/arXiv.astro-ph/0108028},
    urldate = {2024-02-07},
    journal = {arXiv e-prints},
    author = {Phinney, E. S.},
    month = aug,
    year = {2001},
    note = {ADS Bibcode: 2001astro.ph..8028P},
    pages = {astro--ph/0108028},
}

@dataset{le_jeune_2022_7132178,
  author       = {Le Jeune, Maude and
                  Babak, Stanislav},
  title        = {LISA Data Challenge Sangria (LDC2a)},
  month        = oct,
  year         = 2022,
  publisher    = {Zenodo},
  version      = {v2},
  doi          = {10.5281/zenodo.7132178},
  url          = {https://doi.org/10.5281/zenodo.7132178},
}

@misc{criswell_foundation_2026,
    title = {A {Foundation} for {Gravitational}-{Wave} {Population} {Inference} within the {LISA} {Global} {Fit}},
    url = {https://ui.adsabs.harvard.edu/abs/2026arXiv260403390C},
    doi = {10.48550/arXiv.2604.03390},
    urldate = {2026-09-16},
    publisher = {arXiv},
    author = {Criswell, Alexander W. and Banagiri, Sharan and Delfavero, Vera and Bustamante-Rosell, Maria Jose and Taylor, Stephen R. and Rosati, Robert},
    month = apr,
    year = {2026},
    note = {ADS Bibcode: 2026arXiv260403390C},
}

@article{belloni_evidence_2020,
    title = {Evidence for reduced magnetic braking in polars from binary population models},
    volume = {491},
    issn = {0035-8711},
    url = {https://ui.adsabs.harvard.edu/abs/2020MNRAS.491.5717B},
    doi = {10.1093/mnras/stz3413},
    urldate = {2026-09-16},
    journal = {Monthly Notices of the Royal Astronomical Society},
    publisher = {OUP},
    author = {Belloni, Diogo and Schreiber, Matthias R. and Pala, Anna F. and Gänsicke, Boris T. and Zorotovic, Mónica and Rodrigues, Claudia V.},
    month = feb,
    year = {2020},
    note = {ADS Bibcode: 2020MNRAS.491.5717B},
    pages = {5717--5731},
}

@ARTICLE{1999MNRAS.309.1034K,
       author = {{Kolb}, U. and {Baraffe}, I.},
        title = "{Brown dwarfs and the cataclysmic variable period minimum}",
      journal = {mnras},
         year = 1999,
        month = nov,
       volume = {309},
       number = {4},
        pages = {1034-1042},
          doi = {10.1046/j.1365-8711.1999.02926.x},
archivePrefix = {arXiv},
       eprint = {astro-ph/9906448},
 primaryClass = {astro-ph},
       adsurl = {https://ui.adsabs.harvard.edu/abs/1999MNRAS.309.1034K}
}

@ARTICLE{1998PASP..110.1132P,
       author = {{Patterson}, Joseph},
        title = "{Late Evolution of Cataclysmic Variables}",
      journal = {pasp},
         year = 1998,
        month = oct,
       volume = {110},
       number = {752},
        pages = {1132-1147},
          doi = {10.1086/316233},
       adsurl = {https://ui.adsabs.harvard.edu/abs/1998PASP..110.1132P}
}

@article{pretorius_influence_2007,
    title = {The influence of selection effects on the observed cataclysmic variable population: modelling and application to the {Palomar}-{Green} sample},
    volume = {374},
    issn = {0035-8711, 1365-2966},
    shorttitle = {The influence of selection effects on the observed cataclysmic variable population},
    url = {http://arxiv.org/abs/astro-ph/0610278},
    doi = {10.1111/j.1365-2966.2006.11258.x},
    number = {4},
    urldate = {2026-05-15},
    journal = {Monthly Notices of the Royal Astronomical Society},
    author = {Pretorius, Magaretha L. and Knigge, Christian and Kolb, Ulrich},
    month = feb,
    year = {2007},
    note = {arXiv:astro-ph/0610278},
    pages = {1495--1505},
}

@article{lamberts_predicting_2019,
    title = {Predicting the {LISA} white dwarf binary population in the {Milky} {Way} with cosmological simulations},
    volume = {490},
    issn = {0035-8711},
    url = {https://ui.adsabs.harvard.edu/abs/2019MNRAS.490.5888L},
    doi = {10.1093/mnras/stz2834},
    urldate = {2026-07-07},
    journal = {Monthly Notices of the Royal Astronomical Society},
    publisher = {OUP},
    author = {Lamberts, Astrid and Blunt, Sarah and Littenberg, Tyson B. and Garrison-Kimmel, Shea and Kupfer, Thomas and Sanderson, Robyn E.},
    month = dec,
    year = {2019},
    note = {ADS Bibcode: 2019MNRAS.490.5888L},
    pages = {5888--5903},
}

@ARTICLE{2009MNRAS.397.2170G,
       author = {{G{\"a}nsicke}, B.~T. and {Dillon}, M. and {Southworth}, J. and {Thorstensen}, J.~R. and {Rodr{\'\i}guez-Gil}, P. and {Aungwerojwit}, A. and {Marsh}, T.~R. and {Szkody}, P. and {Barros}, S.~C.~C. and {Casares}, J. and {de Martino}, D. and {Groot}, P.~J. and {Hakala}, P. and {Kolb}, U. and {Littlefair}, S.~P. and {Mart{\'\i}nez-Pais}, I.~G. and {Nelemans}, G. and {Schreiber}, M.~R.},
        title = "{SDSS unveils a population of intrinsically faint cataclysmic variables at the minimum orbital period}",
      journal = {mnras},
         year = 2009,
        month = aug,
       volume = {397},
       number = {4},
        pages = {2170-2188},
          doi = {10.1111/j.1365-2966.2009.15126.x},
archivePrefix = {arXiv},
       eprint = {0905.3476},
 primaryClass = {astro-ph.SR},
       adsurl = {https://ui.adsabs.harvard.edu/abs/2009MNRAS.397.2170G}
}

@ARTICLE{1993A&A...271..149K,
       author = {{Kolb}, U.},
        title = "{A model for the intrinsic population of cataclysmic variables}",
      journal = {aap},
         year = 1993,
        month = apr,
       volume = {271},
        pages = {149},
       adsurl = {https://ui.adsabs.harvard.edu/abs/1993A&A...271..149K}
}

@ARTICLE{2001ApJ...550..897H,
       author = {{Howell}, Steve B. and {Nelson}, Lorne A. and {Rappaport}, Saul},
        title = "{An Exploration of the Paradigm for the 2-3 Hour Period Gap in Cataclysmic Variables}",
      journal = {\apj},
         year = 2001,
        month = apr,
       volume = {550},
       number = {2},
        pages = {897-918},
          doi = {10.1086/319776},
archivePrefix = {arXiv},
       eprint = {astro-ph/0005435},
 primaryClass = {astro-ph},
       adsurl = {https://ui.adsabs.harvard.edu/abs/2001ApJ...550..897H}
}

@ARTICLE{2015ApJ...809...80G,
       author = {{Goliasch}, Jonas and {Nelson}, Lorne},
        title = "{Population Synthesis of Cataclysmic Variables. I. Inclusion of Detailed Nuclear Evolution}",
      journal = {\apj},
         year = 2015,
        month = aug,
       volume = {809},
       number = {1},
          eid = {80},
        pages = {80},
          doi = {10.1088/0004-637X/809/1/80},
archivePrefix = {arXiv},
       eprint = {1607.06217},
 primaryClass = {astro-ph.SR},
       adsurl = {https://ui.adsabs.harvard.edu/abs/2015ApJ...809...80G}
}

@ARTICLE{1997CQGra..14.1513S,
       author = {{Schilling}, Roland},
        title = "{Angular and frequency response of LISA}",
      journal = {Classical and Quantum Gravity},
         year = 1997,
        month = jun,
       volume = {14},
       number = {6},
        pages = {1513-1519},
          doi = {10.1088/0264-9381/14/6/020},
       adsurl = {https://ui.adsabs.harvard.edu/abs/1997CQGra..14.1513S}
}

@ARTICLE{2021PhRvD.104d3019K,
       author = {{Karnesis}, Nikolaos and {Babak}, Stanislav and {Pieroni}, Mauro and {Cornish}, Neil and {Littenberg}, Tyson},
        title = "{Characterization of the stochastic signal originating from compact binary populations as measured by LISA}",
      journal = {\prd},
         year = 2021,
        month = aug,
       volume = {104},
       number = {4},
          eid = {043019},
        pages = {043019},
          doi = {10.1103/PhysRevD.104.043019},
archivePrefix = {arXiv},
       eprint = {2103.14598},
 primaryClass = {astro-ph.IM},
       adsurl = {https://ui.adsabs.harvard.edu/abs/2021PhRvD.104d3019K}
}

@ARTICLE{2026PhRvD.113l3006D,
       author = {{De Santi}, Federico and {Santini}, Alessandro and {Toubiana}, Alexandre and {Karnesis}, Nikolaos and {Gerosa}, Davide},
        title = "{Inferring the population properties of Galactic binaries from LISA's stochastic foreground}",
      journal = {\prd},
         year = 2026,
        month = jun,
       volume = {113},
       number = {12},
          eid = {123006},
        pages = {123006},
          doi = {10.1103/9lc5-vgw5},
archivePrefix = {arXiv},
       eprint = {2602.18560},
 primaryClass = {astro-ph.HE},
       adsurl = {https://ui.adsabs.harvard.edu/abs/2026PhRvD.113l3006D}
}

@ARTICLE{2015MNRAS.450...53H,
       author = {{Hopkins}, Philip F.},
        title = "{A new class of accurate, mesh-free hydrodynamic simulation methods}",
      journal = {mnras},
         year = 2015,
        month = jun,
       volume = {450},
       number = {1},
        pages = {53-110},
          doi = {10.1093/mnras/stv195},
archivePrefix = {arXiv},
       eprint = {1409.7395},
 primaryClass = {astro-ph.CO},
       adsurl = {https://ui.adsabs.harvard.edu/abs/2015MNRAS.450...53H}
}

@ARTICLE{2016ApJ...827L..23W,
       author = {{Wetzel}, Andrew R. and {Hopkins}, Philip F. and {Kim}, Ji-hoon and {Faucher-Gigu{\`e}re}, Claude-Andr{\'e} and {Kere{\v{s}}}, Du{\v{s}}an and {Quataert}, Eliot},
        title = "{Reconciling Dwarf Galaxies with {\ensuremath{\Lambda}}CDM Cosmology: Simulating a Realistic Population of Satellites around a Milky Way-mass Galaxy}",
      journal = {apjl},
         year = 2016,
        month = aug,
       volume = {827},
       number = {2},
          eid = {L23},
        pages = {L23},
          doi = {10.3847/2041-8205/827/2/L23},
archivePrefix = {arXiv},
       eprint = {1602.05957},
 primaryClass = {astro-ph.GA},
       adsurl = {https://ui.adsabs.harvard.edu/abs/2016ApJ...827L..23W}
}

@ARTICLE{2018MNRAS.480..800H,
       author = {{Hopkins}, Philip F. and {Wetzel}, Andrew and {Kere{\v{s}}}, Du{\v{s}}an and {Faucher-Gigu{\`e}re}, Claude-Andr{\'e} and {Quataert}, Eliot and {Boylan-Kolchin}, Michael and {Murray}, Norman and {Hayward}, Christopher C. and {Garrison-Kimmel}, Shea and {Hummels}, Cameron and {Feldmann}, Robert and {Torrey}, Paul and {Ma}, Xiangcheng and {Angl{\'e}s-Alc{\'a}zar}, Daniel and {Su}, Kung-Yi and {Orr}, Matthew and {Schmitz}, Denise and {Escala}, Ivanna and {Sanderson}, Robyn and {Grudi{\'c}}, Michael Y. and {Hafen}, Zachary and {Kim}, Ji-Hoon and {Fitts}, Alex and {Bullock}, James S. and {Wheeler}, Coral and {Chan}, T.~K. and {Elbert}, Oliver D. and {Narayanan}, Desika},
        title = "{FIRE-2 simulations: physics versus numerics in galaxy formation}",
      journal = {mnras},
         year = 2018,
        month = oct,
       volume = {480},
       number = {1},
        pages = {800-863},
          doi = {10.1093/mnras/sty1690},
archivePrefix = {arXiv},
       eprint = {1702.06148},
 primaryClass = {astro-ph.GA},
       adsurl = {https://ui.adsabs.harvard.edu/abs/2018MNRAS.480..800H}
}

@article{phan2019composable,
    title   = {Composable Effects for Flexible and Accelerated Probabilistic Programming in NumPyro},
    author  = {Phan, Du and Pradhan, Neeraj and Jankowiak, Martin},
    journal = {arXiv preprint arXiv:1912.11554},
    year    = {2019}
}

@article{bingham2019pyro,
    author  = {Eli Bingham and
               Jonathan P. Chen and
               Martin Jankowiak and
               Fritz Obermeyer and
               Neeraj Pradhan and
               Theofanis Karaletsos and
               Rohit Singh and
               Paul A. Szerlip and
               Paul Horsfall and
               Noah D. Goodman},
    title   = {Pyro: Deep Universal Probabilistic Programming},
    journal = {J. Mach. Learn. Res.},
    volume  = {20},
    pages   = {28:1--28:6},
    year    = {2019},
    url     = {http://jmlr.org/papers/v20/18-403.html}
}

@software{sergey_koposov_2025_17268284,
  author       = {Sergey Koposov and
                  Josh Speagle and
                  Kyle Barbary and
                  Gregory Ashton and
                  Ed Bennett and
                  Johannes Buchner and
                  Carl Scheffler and
                  Colm Talbot and
                  Ben Cook and
                  James Guillochon and
                  Patricio Cubillos and
                  Andrés Asensio Ramos and
                  Matthieu Dartiailh and
                  Ilya and
                  Erik Tollerud and
                  Dustin Lang and
                  Ben Johnson and
                  jtmendel and
                  Edward Higson and
                  Thomas Vandal and
                  Tansu Daylan and
                  Ruth Angus and
                  patelR and
                  Phillip Cargile and
                  Patrick Sheehan and
                  Matt Pitkin and
                  Matthew Kirk and
                  Lu Xu and
                  Joel Leja and
                  joezuntz},
  title        = {joshspeagle/dynesty: v3.0.0},
  month        = oct,
  year         = 2025,
  publisher    = {Zenodo},
  version      = {v3.0.0},
  doi          = {10.5281/zenodo.17268284},
  url          = {https://doi.org/10.5281/zenodo.17268284},
  swhid        = {swh:1:dir:b3340f6aaa931bae6c6bcff6e7ddeb89ca508a15
                   ;origin=https://doi.org/10.5281/zenodo.3348367;vis
                   it=swh:1:snp:fa92422c27d5611b107216e9b766a871bb43b
                   bee;anchor=swh:1:rel:94af3a8ef6e50d997258e152c3833
                   3550bd7463f;path=joshspeagle-dynesty-217bc94
                  },
}

@ARTICLE{2020MNRAS.493.3132S,
       author = {{Speagle}, Joshua S.},
        title = "{DYNESTY: a dynamic nested sampling package for estimating Bayesian posteriors and evidences}",
      journal = {mnras},
         year = 2020,
        month = apr,
       volume = {493},
       number = {3},
        pages = {3132-3158},
          doi = {10.1093/mnras/staa278},
archivePrefix = {arXiv},
       eprint = {1904.02180},
 primaryClass = {astro-ph.IM},
       adsurl = {https://ui.adsabs.harvard.edu/abs/2020MNRAS.493.3132S}
}

@INPROCEEDINGS{2004AIPC..735..395S,
       author = {{Skilling}, John},
        title = "{Nested Sampling}",
    booktitle = {Bayesian Inference and Maximum Entropy Methods in Science and Engineering: 24th International Workshop on Bayesian Inference and Maximum Entropy Methods in Science and Engineering},
         year = 2004,
       editor = {{Fischer}, Rainer and {Preuss}, Roland and {Toussaint}, Udo Von},
       series = {American Institute of Physics Conference Series},
       volume = {735},
        month = nov,
    publisher = {AIP},
        pages = {395-405},
          doi = {10.1063/1.1835238},
       adsurl = {https://ui.adsabs.harvard.edu/abs/2004AIPC..735..395S}
}

@article{10.1214/06-BA127,
author = {John Skilling},
title = {{Nested sampling for general Bayesian computation}},
volume = {1},
journal = {Bayesian Analysis},
number = {4},
publisher = {International Society for Bayesian Analysis},
pages = {833 -- 859},
year = {2006},
doi = {10.1214/06-BA127},
URL = {https://doi.org/10.1214/06-BA127}
}

@Article{Hunter:2007,
  Author    = {Hunter, J. D.},
  Title     = {Matplotlib: A 2D graphics environment},
  Journal   = {Computing in Science \& Engineering},
  Volume    = {9},
  Number    = {3},
  Pages     = {90--95},
  publisher = {IEEE COMPUTER SOC},
  doi       = {10.1109/MCSE.2007.55},
  year      = 2007
}

@article{corner,
  doi = {10.21105/joss.00024},
  url = {https://doi.org/10.21105/joss.00024},
  year  = {2016},
  month = {jun},
  publisher = {The Open Journal},
  volume = {1},
  number = {2},
  pages = {24},
  author = {Daniel Foreman-Mackey},
  title = {corner.py: Scatterplot matrices in Python},
  journal = {The Journal of Open Source Software}
}

@inproceedings{soton403913,
       booktitle = {Positioning and Power in Academic Publishing: Players, Agents and Agendas},
          editor = {Fernando Loizides and Birgit Scmidt},
           title = {Jupyter Notebooks ? a publishing format for reproducible computational workflows},
          author = {Thomas Kluyver and Benjamin Ragan-Kelley and Fernando P{\'e}rez and Brian Granger and Matthias Bussonnier and Jonathan Frederic and Kyle Kelley and Jessica Hamrick and Jason Grout and Sylvain Corlay and Paul Ivanov and Dami{\'a}n Avila and Safia Abdalla and Carol Willing and  Jupyter development team},
       publisher = {IOS Press},
            year = {2016},
           pages = {87--90},
             url = {https://eprints.soton.ac.uk/403913/}
}

@Article{         harris2020array,
 title         = {Array programming with {NumPy}},
 author        = {Charles R. Harris and K. Jarrod Millman and St{\'{e}}fan J.
                 van der Walt and Ralf Gommers and Pauli Virtanen and David
                 Cournapeau and Eric Wieser and Julian Taylor and Sebastian
                 Berg and Nathaniel J. Smith and Robert Kern and Matti Picus
                 and Stephan Hoyer and Marten H. van Kerkwijk and Matthew
                 Brett and Allan Haldane and Jaime Fern{\'{a}}ndez del
                 R{\'{i}}o and Mark Wiebe and Pearu Peterson and Pierre
                 G{\'{e}}rard-Marchant and Kevin Sheppard and Tyler Reddy and
                 Warren Weckesser and Hameer Abbasi and Christoph Gohlke and
                 Travis E. Oliphant},
 year          = {2020},
 month         = sep,
 journal       = {Nature},
 volume        = {585},
 number        = {7825},
 pages         = {357--362},
 doi           = {10.1038/s41586-020-2649-2},
 publisher     = {Springer Science and Business Media {LLC}},
 url           = {https://doi.org/10.1038/s41586-020-2649-2}
}

@ARTICLE{2020SciPy-NMeth,
  author  = {Virtanen, Pauli and Gommers, Ralf and Oliphant, Travis E. and
            Haberland, Matt and Reddy, Tyler and Cournapeau, David and
            Burovski, Evgeni and Peterson, Pearu and Weckesser, Warren and
            Bright, Jonathan and {van der Walt}, St{\'e}fan J. and
            Brett, Matthew and Wilson, Joshua and Millman, K. Jarrod and
            Mayorov, Nikolay and Nelson, Andrew R. J. and Jones, Eric and
            Kern, Robert and Larson, Eric and Carey, C J and
            Polat, {\.I}lhan and Feng, Yu and Moore, Eric W. and
            {VanderPlas}, Jake and Laxalde, Denis and Perktold, Josef and
            Cimrman, Robert and Henriksen, Ian and Quintero, E. A. and
            Harris, Charles R. and Archibald, Anne M. and
            Ribeiro, Ant{\^o}nio H. and Pedregosa, Fabian and
            {van Mulbregt}, Paul and {SciPy 1.0 Contributors}},
  title   = {{{SciPy} 1.0: Fundamental Algorithms for Scientific
            Computing in Python}},
  journal = {Nature Methods},
  year    = {2020},
  volume  = {17},
  pages   = {261--272},
  adsurl  = {https://rdcu.be/b08Wh},
  doi     = {10.1038/s41592-019-0686-2},
}

@software{jax2018github,
    author = {James Bradbury and Roy Frostig and Peter Hawkins and Matthew James Johnson and Yash Katariya and Chris Leary and Dougal Maclaurin and George Necula and Adam Paszke and Jake Vander{P}las and Skye Wanderman-{M}ilne and Qiao Zhang},
    title = {{JAX}: composable transformations of {P}ython+{N}um{P}y programs},
    url = {http://github.com/jax-ml/jax},
    version = {0.3.13},
    year = {2018},
}

\end{document}